\documentclass{aa}

\usepackage{graphicx}
\usepackage{txfonts}
\usepackage{natbib, twoopt}
\usepackage{subfig}
\usepackage{color}
\usepackage{mathrsfs}
\usepackage{xspace}
\usepackage{rotating}
\usepackage{lscape}
\usepackage{amssymb}
\usepackage{longtable}
\usepackage{pifont}
\usepackage{multirow}
\usepackage{adjustbox}
\usepackage{lineno}
\usepackage{gensymb}
\usepackage{mathtools}
\usepackage[plainpages=False,bookmarks=True,colorlinks=True,citecolor=blue,linkcolor=blue,urlcolor=magenta,breaklinks=True,pagebackref=False]{hyperref}

\newcommand{\Nsn}{145\xspace}
\newcommand{\NsnCl}{11\xspace}

\newcommand{\wrmsClusDoppler}{$0.130 \pm 0.038$\xspace}
\newcommand{\wrmsClusNoDoppler}{$0.137 \pm 0.036$ \xspace}

\newcommand{\wrmsOutClusDoppler}{$0.151 \pm 0.010$\xspace}

\newcommand{\pvalueNsim}{$10^6$\xspace}

\newcommand{\pvalueCluster}{$5.9\times10^{-4}$\xspace}
\newcommand{\pvalueClusterwithoutSNX}{$6.6\times10^{-2}$\xspace}

\newcommand{\pearsonCluster}{$\rho = 0.9 \pm 0.1$\xspace}
\newcommand{\pearsonClusterwithoutSNX}{$\rho = 0.5 \pm 0.3 $ \xspace}

\newcommand{\pearsonsigCluster}{$3.58 \ \sigma$\xspace}
\newcommand{\pearsonsigClusterwithoutSNX}{$1.34 \ \sigma$\xspace}

\newcommand{\pvalueDispCluster}{$3.8\times10^{-1}$\xspace}

\newcommand{\sne}{SNe~Ia}
\newcommand{\zobs}{\ensuremath{z_{obs}}}
\newcommand{\zc}{\ensuremath{z_{c}}}

\newcommand{\zp}{\ensuremath{z_{p}}}

\newcommand{\rdeuxcent}{\ensuremath{R_{200}}}
\newcommand{\mdeuxcent}{\ensuremath{M_{200}}}

\newcommand{\sigv}{\ensuremath{\sigma_V}}

\newcommand{\mbstar}{\ensuremath{m_B^*}}

\newcommand{\wrms}{\ensuremath{wRMS}}

\begin{document} 
 \graphicspath{{figures/}} 

   \title{Correcting for peculiar velocities of Type Ia Supernovae in clusters of galaxies}   
   \titlerunning{Type Ia Supernovae \& Galaxy clusters}
   \authorrunning{P.-F. L\'eget \& SNfactory}
   \date{Received 1 March 2018 / Accepted 6 April 2018}
\author{P.-F.~L\'eget \inst{\ref{lpc},\ref{kipac}}
\and   M.~V. Pruzhinskaya\inst{\ref{lpc},\ref{msu}}
\and   A.~Ciulli \inst{\ref{lpc}}
\and   E.~Gangler \inst{\ref{lpc}}
\and   G.~Aldering \inst{\ref{lbnl}}
\and   P.~Antilogus \inst{\ref{lpnhe}}
\and   C.~Aragon \inst{\ref{lbnl}}
\and   S.~Bailey \inst{\ref{lbnl}}
\and   C.~Baltay \inst{\ref{yale}} 
\and   K.~Barbary \inst{\ref{lbnl}} 
\and   S.~Bongard \inst{\ref{lpnhe}}
\and   K.~Boone \inst{\ref{lbnl},\ref{ucb}}
\and   C.~Buton \inst{\ref{ipnl}}
\and   M.~Childress \inst{\ref{southampton}}
\and   N.~Chotard \inst{\ref{ipnl}}
\and   Y.~Copin \inst{\ref{ipnl}}
\and   S.~Dixon \inst{\ref{lbnl}}
\and   P.~Fagrelius \inst{\ref{lbnl},\ref{ucb}}
\and   U.~Feindt \inst{\ref{okc}}
\and   D.~Fouchez \inst{\ref{cppm}}
\and   P.~Gris \inst{\ref{lpc}}
\and   B.~Hayden \inst{\ref{lbnl}}
\and   W.~Hillebrandt \inst{\ref{garching}}
\and   D.~A. Howell \inst{\ref{lcogt},\ref{ucsb}} 
\and   A.~Kim \inst{\ref{lbnl}}
\and   M.~Kowalski \inst{\ref{berlin},\ref{desy}}
\and   D.~Kuesters \inst{\ref{berlin}}
\and   S.~Lombardo \inst{\ref{berlin}}
\and   Q.~Lin \inst{\ref{china}}
\and   J.~Nordin \inst{\ref{berlin}}
\and   R.~Pain \inst{\ref{lpnhe}}
\and   E.~Pecontal \inst{\ref{cral}}
\and   R.~Pereira \inst{\ref{ipnl}}
\and   S.~Perlmutter \inst{\ref{lbnl},\ref{ucb}}
\and   D.~Rabinowitz \inst{\ref{yale}} 
\and   M.~Rigault \inst{\ref{lpc}} 
\and   K.~Runge \inst{\ref{lbnl}} 
\and   D.~Rubin \inst{\ref{lbnl},\ref{stsci}}
\and   C.~Saunders \inst{\ref{lpnhe}}
\and   L.-P.~Says \inst{\ref{lpc}}   
\and   G.~Smadja \inst{\ref{ipnl}} 
\and   C.~Sofiatti \inst{\ref{lbnl},\ref{ucb}}
\and   N.~Suzuki \inst{\ref{lbnl},\ref{ipmu}}
\and   S.~Taubenberger \inst{\ref{garching},\ref{eso}}
\and   C.~Tao \inst{\ref{cppm},\ref{china}}
\and   R.~C.~Thomas \inst{\ref{nersc}}\\
\textsc{The Nearby Supernova Factory}
}

\institute{\tiny
Universit\'e Clermont Auvergne, CNRS/IN2P3, Laboratoire de Physique de Clermont, F-63000 Clermont-Ferrand, France. \label{lpc}
\and
    Kavli Institute for Particle Astrophysics and Cosmology,
    Department of Physics, Stanford University, 
    Stanford, CA 94305 \label{kipac}
\and 
    Lomonosov Moscow State University, Sternberg Astronomical Institute, Universitetsky pr.~13, Moscow 119234, Russia \label{msu}
\and    
    Physics Division, Lawrence Berkeley National Laboratory, 
    1 Cyclotron Road, Berkeley, CA, 94720 \label{lbnl}
\and 
    Laboratoire de Physique Nucl\'eaire et des Hautes \'Energies,
    Universit\'e Pierre et Marie Curie Paris 6, Universit\'e Paris Diderot Paris 7, CNRS-IN2P3, 
    4 place Jussieu, 75252 Paris Cedex 05, France \label{lpnhe}
\and 
    Department of Physics, Yale University, 
    New Haven, CT, 06250-8121 \label{yale}
\and 
    Department of Physics, University of California Berkeley,
    366 LeConte Hall MC 7300, Berkeley, CA, 94720-7300 \label{ucb}
\and 
    Universit\'e de Lyon, F-69622, Lyon, France ; Universit\'e de Lyon 1, Villeurbanne ; 
    CNRS/IN2P3, Institut de Physique Nucl\'eaire de Lyon. \label{ipnl}
\and 
    Department of Physics and Astronomy, University of Southampton,
    Southampton, Hampshire, SO17 1BJ, UK \label{southampton} 
\and 
    The Oskar Klein Centre, Department of Physics, AlbaNova, Stockholm
    University, SE-106 91 Stockholm, Sweden \label{okc}
\and 
    Aix Marseille Universit\'e, CNRS/IN2P3, CPPM UMR 7346, 13288,
    Marseille, France \label{cppm}
\and
    Max-Planck Institut f\"ur Astrophysik, Karl-Schwarzschild-Str. 1,
    85748 Garching, Germany \label{garching}
\and 
    Las Cumbres Observatory Global Telescope Network, 6740 Cortona
    Dr., Suite 102 Goleta, Ca 93117 \label{lcogt} 
\and 
    Department of Physics, University of California, Santa Barbara, CA
    93106-9530, USA \label{ucsb} 
\and
    Institut fur Physik, Humboldt-Universitat zu Berlin,
    Newtonstr. 15, 12489 Berlin \label{berlin} 
\and
    Deutsches Elektronen-Synchrotron, D-15735 Zeuthen, Germany \label{desy}
\and 
    Tsinghua Center for Astrophysics, Tsinghua University, Beijing
    100084, China \label{china} 
\and 
    Centre de Recherche Astronomique de Lyon, Universit\'e Lyon 1,
    9 Avenue Charles Andr\'e, 69561 Saint Genis Laval, France \label{cral}
\and
    Space Telescope Science Institute, 3700 San Martin Drive,
    Baltimore, MD 21218 \label{stsci}
\and
    European Southern Observatory, Karl-Schwarzschild-Str. 2, 85748
    Garching, Germany \label{eso}
\and
    Computational Cosmology Center, Computational Research Division, Lawrence Berkeley National Laboratory, 
    1 Cyclotron Road MS 50B-4206, Berkeley, CA, 94720 \label{nersc}
\and
    Kavli Institute for the Physics and Mathematics of the Universe,
    University of Tokyo, 5-1-5 Kashiwanoha, Kashiwa, Chiba, 277-8583, Japan \label{ipmu}
}

  \abstract
   {Type Ia Supernovae (SNe~Ia) are widely used to measure the expansion of the Universe. To perform such measurements the luminosity and cosmological redshift ($z$) of the SNe~Ia have to be determined. The uncertainty on $z$ includes an unknown peculiar velocity, which can be very large for SNe~Ia in the virialized cores of massive clusters.}
   {We determine which SNe~Ia exploded in galaxy clusters. We then study how the correction for peculiar velocities of host galaxies inside the clusters improves the Hubble residuals.}
   {Using \Nsn SNe~Ia from the Nearby Supernova Factory we found \NsnCl candidates for membership in clusters. To estimate the redshift of a cluster we applied the bi-weight technique. Then, we use the galaxy cluster redshift instead of the host galaxy redshift to construct the Hubble diagram.}
   {For SNe~Ia inside galaxy clusters the dispersion around the Hubble diagram when peculiar velocities are taken into account is smaller in comparison with a case without peculiar velocity correction, with a \wrms=\wrmsClusDoppler mag instead of \mbox{\wrms=\wrmsClusNoDoppler} mag. The significance of this improvement is \pearsonsigCluster. If we remove the very nearby Virgo cluster member SN2006X ($z<0.01$) from the analysis, the significance decreases to \pearsonsigClusterwithoutSNX. The peculiar velocity correction is found to be highest for the SNe~Ia hosted by blue spiral galaxies, with high local specific star formation rate and smaller stellar mass, seemingly counter to what might be expected given the heavy concentration of old, massive elliptical galaxies in clusters.}
   {As expected, the Hubble residuals of SNe~Ia associated with massive galaxy clusters improve when the cluster redshift is taken as the cosmological redshift of the SN. This fact has to be taken into account in future cosmological analyses in order to achieve higher accuracy for cosmological redshift measurements. Here we provide an approach to do so.}
   \maketitle

\section{Introduction}

Type Ia Supernovae (SNe~Ia) are excellent distance indicators. Observations of distant SNe~Ia led to the discovery of the accelerating expansion of the Universe~(\citealt{1998Natur.391...51P,Perlmutter99}, \citealt{Riess98}, \citealt{1998ApJ...507...46S}). The most recent analysis of SNe~Ia indicates that for a flat $\Lambda$CDM cosmology, our Universe is accelerating, with $\Omega_\Lambda = 0.705\pm0.034$ (\citealt{Betoule14,2017arXiv171000845S}).

Cosmological parameters are estimated from the ``luminosity distance-redshift'' relation of \sne, using the Hubble diagram. Generally, particular attention is paid to standardization of \sne, i.e. to increase of the accuracy of luminosity distance determinations~\citep{1974PhDT.........7R,1977SvA....21..675P,1984SvA....28..658P, Phillips93, 1996AJ....112.2391H, Phillips99, Riess96, Perlmutter97, Perlmutter99, Wang03, Guy05, Guy07, Jha07, Bailey09, Wang09b, kelly10, Sullivan10, Chotard11, Blondin12a, Rigault13, Kim13, Boone15, Sasdelli16, LegetPhD, Saunders16}. The uncertainty on the redshift is quite often considered negligible. The redshift used in ``luminosity distance-redshift'' relation is due to the expansion of the Universe assuming Friedman-Lemaitre-Robertson-Walker metric, i.e. the motion within the reference frame defined by the cosmic microwave background radiation (CMB). We will refer to this as a cosmological redshift ($\zc$). In fact, the redshift observed on the Earth ($\zobs$) also includes the contribution from the Doppler effect induced by radial peculiar velocities ($\zp$):

\begin{equation}
(1+\zobs)=(1+\zc)(1+\zp)
\label{composition_redshifts}
\end{equation}

At low redshift, and for low velocities compared to the speed of light in vacuum, the following approximation can be used: 

 \begin{equation}
\zobs=\zc+\zp
\label{approx_compos_redsh} 
\end{equation}

The component of the redshift due to peculiar velocities includes the Earth's rotational and orbital motions, the Solar orbit within the Galaxy, peculiar motion of the Galaxy within the Local Group, ``infall'' of the Local Group toward the center of the Local Supercluster, etc. It is well known that peculiar velocities of \sne~introduce additional errors to the Hubble diagram and therefore have an impact on the estimation of cosmological parameters~\citep{Cooray2006,Hui2006,Davis11,Habibi14}. To minimize the influence of poorly constrained peculiar velocities, in some cosmological analyses all SNe~Ia  with $z<0.015$ are removed from the Hubble diagram fitting and a 300--400 km~s$^{-1}$ peculiar velocity dispersion is added in quadrature to the redshift uncertainty~\citep{Astier06,Wood-Vasey07,Amanullah10}. In particular, this is the approach taken for the cosmology analysis using Union 2.1~\citep{Suzuki2012}.
Another way to apply the peculiar velocity correction is to measure the local velocity field assuming linear perturbation theory and then correct each supernova redshift~\citep{Hudson2004}.  
\cite{Willick98} estimated the accuracy of this method to be $\sim$100 km~s$^{-1}$, \cite{Riess97} adopted the value of 200 km~s$^{-1}$, \cite{Conley2011} used 150 km~s$^{-1}$. This approach was used in the Joint Light-Curve Analysis (JLA;~\citealt{Betoule14}). 
However, it has been shown that the systematic uncertainty on $w$, the dark energy equation of state parameter, of different flow models is at the level of $\pm$0.04~\citep{Neill2007}.

\begin{figure}
\begin{center}
\includegraphics[scale=0.45]{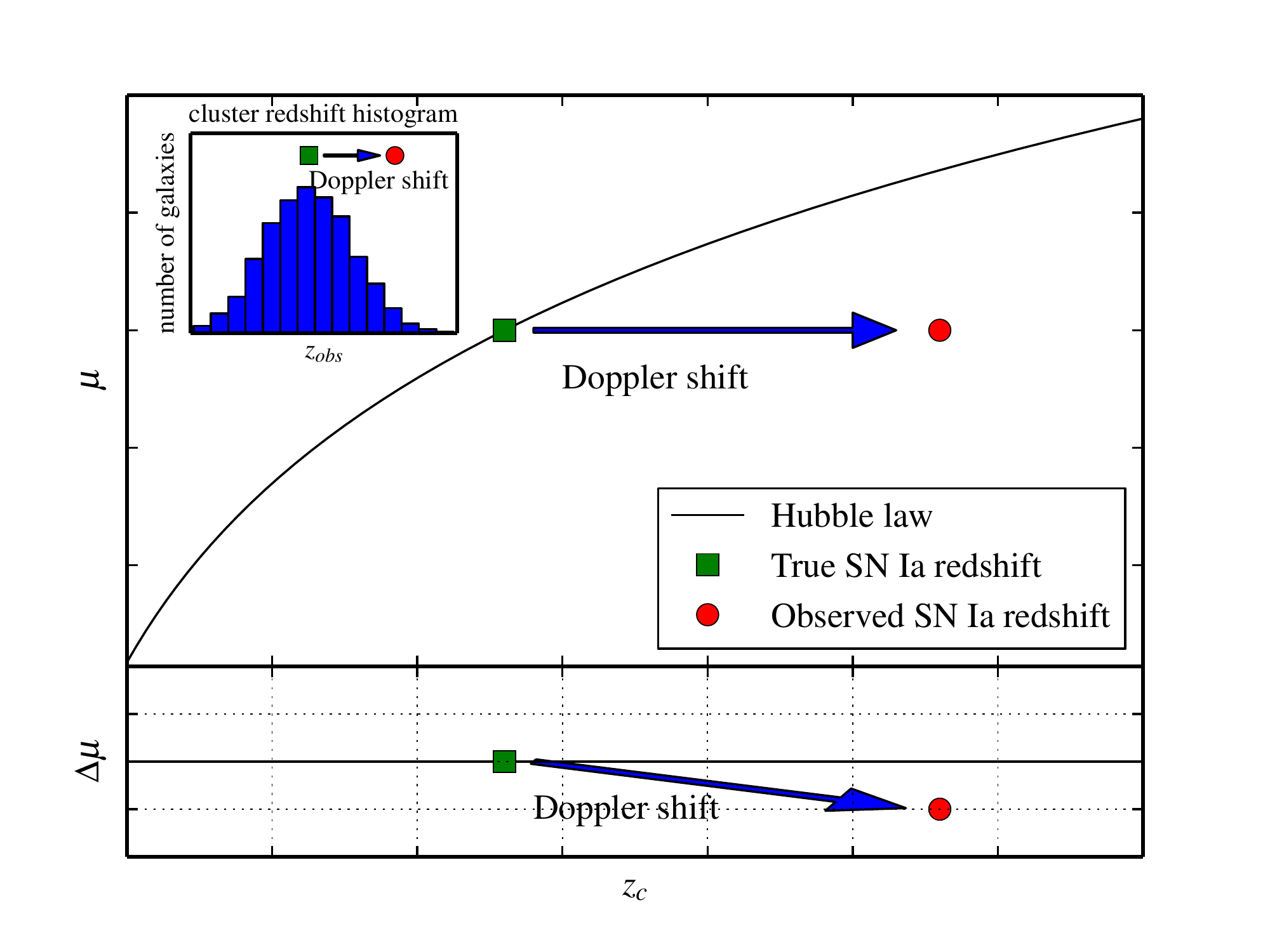}
\caption{This Hubble diagram demonstrates how large peculiar velocity can affect the measurements of the expansion history of the Universe. The inset plot is a typical velocity distribution of galaxies inside a cluster.} 
\label{expl}
\end{center}
\end{figure}

It has nonetheless been observed that velocity dispersions can exceed 1000 km~s$^{-1}$~in galaxy clusters \citep{Ruel2014}. For example, in the Coma cluster, a large cluster of galaxies that contains more than 1000 members, the velocity dispersion is $\sigma_V= 1038$ km~s$^{-1}$ (\citealt{Colless1996}).  The dispersion inside the cluster can be much greater than that usually assumed in cosmological analyses and therefore can seriously affect the redshift measurements (see~Fig.~\ref{expl}). Moreover, within a cluster, the perturbations are no longer linear, and therefore can not be corrected using the smoothed velocity field. Assuming a linear Hubble flow, we can transform the dispersion due to peculiar velocities into a magnitude error: 

\begin{equation}
\label{erreur_vitesse_prope}
\sigma_m =\frac{5 \, \sigma_V}{c z \ln (10)}.
\end{equation}

\begin{figure}
\begin{center}
\includegraphics[scale=0.27]{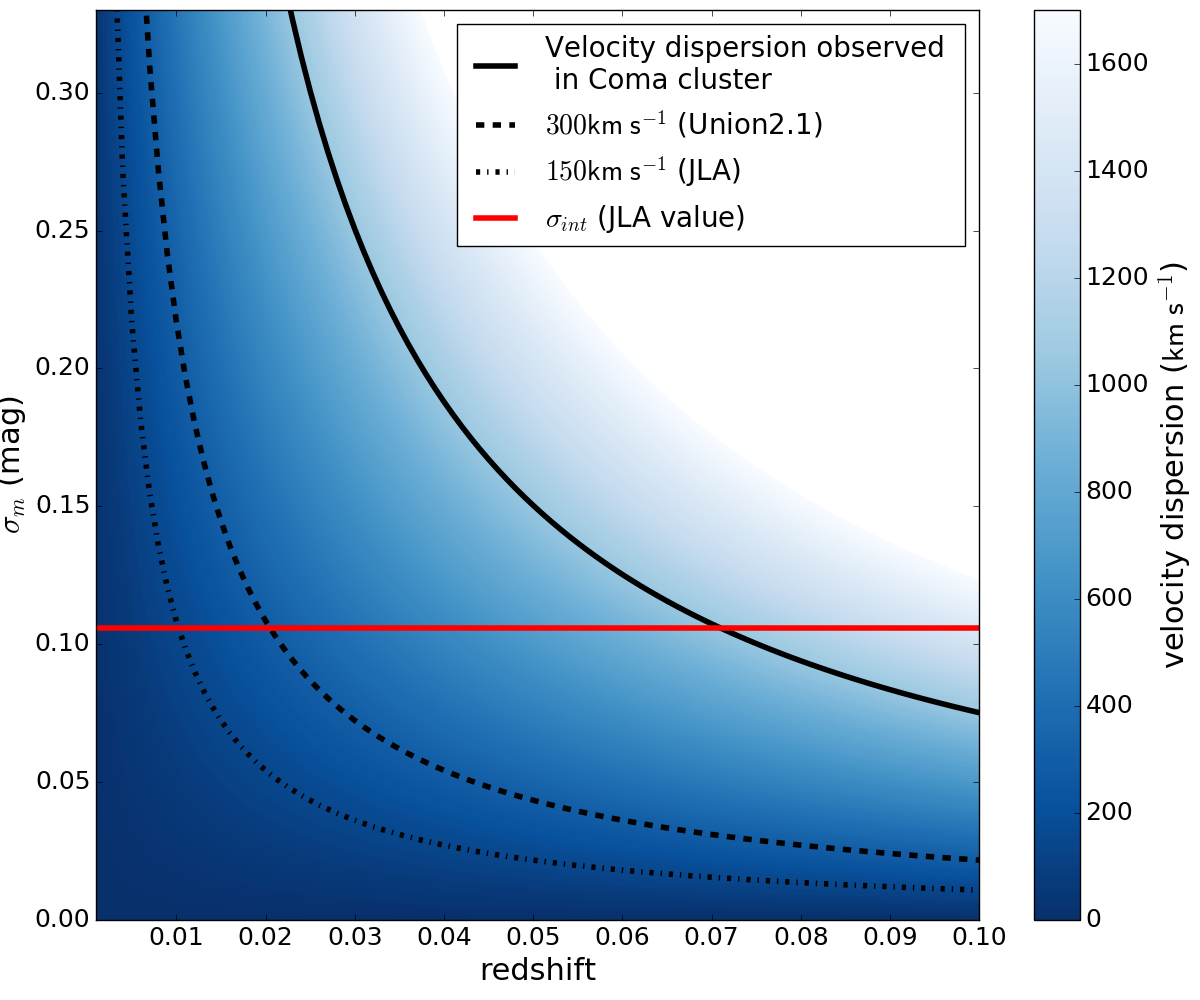}
\caption{Redshift uncertainties (in magnitude units) due to different levels of peculiar velocities, as a function of the cosmological redshift. The solid black line corresponds to the Coma cluster velocity dispersion; the dashed and dash-dotted lines correspond to 300 km~s$^{-1}$~and 150 km~s$^{-1}$, respectively. The red line shows the intrinsic dispersion of SNe~Ia on the Hubble diagram found for the JLA sample~\citep{Betoule14}.}
\label{redshift_distancelum}
\end{center}
\end{figure}

Calculations using Eq.~\ref{erreur_vitesse_prope} show that for the low redshift region ($z < 0.05$) this error is higher than the 150~km~s$^{-1}$ and 300 km~s$^{-1}$ that is usually assumed and is two times larger than the intrinsic dispersion of SNe~Ia around the Hubble diagram~(Fig.~\ref{redshift_distancelum}). This means that standard methods to take into account peculiar velocities do not work for galaxies inside clusters, and another more accurate method needs to be developed for these special cases.

For a SN in a cluster it is possible to estimate $\zc$ more accurately using the host galaxy cluster redshift ($z^{\,cl}$) instead of the host redshift\footnote{Hereafter, we refer to this procedure as peculiar velocity correction.} ($z^{\,host}$). The mean cluster redshift is not affected by virialization within a cluster. 
Of course clusters also have peculiar velocities which can sometimes manifest themselves as cluster merging, for example, the Bullet clusters \citep{Clowe06}. However, clusters have much smaller peculiar velocities than the galaxies within them (i.e. $\sim$300~km~s$^{-1}$; \citealt{Bahcall1996,Dale1999,Masters2006}).

The fact that there is additional velocity dispersion of galaxies inside the clusters that should be taken into account has been known for a long time. Indeed, the distance measurements are degenerate in terms of redshift due to the presence of galaxy clusters and this is accounted for when Tully-Fisher method \citep{Tully77} is applied to measure distances. This problem is known as the triple value problem, which is the fact that for a given distance one can get three different values of redshift due to the presence of a cluster (see for example \citealt{Tonry81,Tully84,Blakeslee99,Radburn-Smith04,Karachentsev14}). To account for the peculiar velocities of galaxies in clusters \cite{Blakeslee99} proposed several alternative approaches. The first is to keep using the individual galaxies' velocities but to add extra variance in quadrature for the clusters according to $\sigma_{cl}(r)=\sigma_0/[1+(r/r_0)^2]^{1/2}$, where $\sigma_0=700 ~(400)$ km~s$^{-1}$ and $r_0=2~(1)$ Mpc for Virgo (Fornax). The second approach is to use a fixed velocity error and to remove the virial dispersion by assigning galaxies their group-averaged velocities. Nevertheless, peculiar velocity correction within galaxy clusters has received little attention in SN~Ia studies, with the exceptions of \citealt{Feindt13} and \citealt{2017arXiv170700715D}. The redshift correction induced by galaxy clusters is mentioned only briefly in those analyses, as their objectives were to measure the bulk flow with SNe~Ia~\citep{Feindt13} and the Hubble constant~\citep{2017arXiv170700715D}. However, at low redshifts this correction is necessary, which is why we focus on it here.

In this paper we identify \sne~that appear to reside in known clusters of galaxies. We then estimate the impact of their peculiar velocities by replacing the host redshift by the cluster redshift. As our parent sample we use \Nsn \sne~observed by the Nearby Supernova Factory (\textsc{SNfactory}), a project devoted to the study of \sne~in the nearby Hubble flow ($0.02 < z < 0.08$;~\citealt{Aldering02}). We then compare the Hubble residuals (HRs) for SNe~Ia in galaxy clusters before and after peculiar velocity correction.
 
The paper is organized as follows: in Sect.~\ref{data} the \textsc{SNfactory} dataset is described. In Sect.~\ref{hostdata} the host clusters data as well as the matching with SNe~Ia are presented. In Sect.~\ref{impactHD} we introduce the peculiar velocity correction and study how it affects the HRs. We discuss the robustness of our results and the properties of SNe~Ia in galaxy clusters in Sect.~\ref{discussion}. Finally, the conclusions of this study are given in Sect.~\ref{conclusions}.

Throughout this paper, we assume a flat $\Lambda$CDM cosmology with $\Omega_\Lambda=0.7$, $\Omega_m=0.3$, and $H_0 = 70$ km~s$^{-1}$Mpc$^{-1}$. Varying these assumptions has negligible impact on our results due to the low redshifts of our SNe~Ia and the fact that $H_0$ is simply absorbed into the Hubble diagram zero point.

\section{Nearby Supernova factory data}
\label{data}
This analysis is based on \Nsn SNe~Ia obtained by the \textsc{SNfactory} collaboration between 2004 and 2009  with the SuperNova Integral Field Spectrograph (SNIFS; \citealt{Aldering02}, \citealt{Lantz04}) installed on the University of Hawaii 2.2-m telescope (Mauna Kea). SNIFS is a fully integrated instrument optimized for semi-automated observations of point sources on a structured background over an extended optical window at moderate spectral resolution. SNIFS has a fully-filled $6.4''\times6.4''$ spectroscopic field-of-view subdivided into a grid of $15\times15$ contiguous square spatial elements (spaxels). The dual-channel spectrograph simultaneously covers 3200--5200 \AA{} (B-channel) and 5100--10000 \AA{} (R-channel) with 2.8 and 3.2 \AA~resolution, respectively. The data reduction of the x, y, $\lambda$ data cubes was summarized by \cite{Aldering06} and updated in Sect.~2.1 of  \cite{Scalzo10}. A preview of the flux calibration is developed in Sect. 2.2 of \cite{Pereira13}, based on the atmospheric extinction derived in \cite{Buton13}, and the host subtraction is described in \cite{Bongard11}. For every supernova followed, the \textsc{SNfactory} creates a spectro-photometric time series composed of $\sim$13 epochs on average, with the first spectrum taken before maximum light in $B$-band \citep{Bailey09,Chotard11}. In addition, observations are obtained at the supernova location at least one year after the explosion to serve as a final reference to enable the subtraction of the underlying host. The host galaxy redshifts of the \textsc{SNfactory} SNe~Ia are given in \citealt{2013ApJ...770..107C}. The sample of \Nsn SNe~Ia contains those objects through 2009 having good final references and properly measured light-curve parameters, including quality cuts suggested by \cite{Guy10}. 

The nearby supernova search is more complicated than the search for distant SNe~Ia because, to probe the same volume, it is necessary to sweep a much larger sky field. Rather than targeting high-density galaxy fields that could potentially bias the survey, at the beginning of the \textsc{SNfactory} experiment (2004--2008), SNe~Ia were discovered with the 1.2-m telescope at the Mount Palomar Observatory \citep{Rabinowitz03} in a non-targeted mode, by surveying about 500 square degrees of sky every night. In all $\sim$20000 square degrees were monitored over the course of a year. \textsc{SNfactory} performed the follow-up observations of a few SNe~Ia discovered by the Palomar Transient Factory \citep{Law09} which also were found in a non-targeted search. We chose to examine this sample, despite it being only 20\% of all nearby cosmologically useful SNe~Ia, in order to use a homogeneous dataset primarily from a blind SN~Ia search to avoid any bias due to the survey strategy.  However, 22 SNe~Ia in the sample were not discovered by these research programs but by amateur astronomers or specific surveys in clusters of galaxies. In particular, SN2007nq which will be identified as being in a cluster, comes from a specific search within clusters of galaxies \citep{2007CBET.1106....1Q}; SN2006X as well as SN2009hi which were also identified as being in clusters, come from targeted searches~\citep{2006IAUC.8667....1P,2009CBET.1872....1N}.

As mentioned above, SN2006X is located in the Virgo cluster and is a highly reddened SN~Ia, with a SALT2 color of $C=1.2$. This SN~Ia would not be kept for a classical cosmological analysis, but since here we are only interested in the effects of peculiar velocities, we have kept it in the analysis.

\section{Host clusters data}
\label{hostdata}
In this section we will describe how we selected the cluster candidates for associations with \textsc{SNfactory} SNe~(Sect.~\ref{clusterselection}). We will then present our technique for calculating the cluster redshift and its error~(Sect.~\ref{section_redshfit_measurement}). Our final list of associations appears in Sect.~\ref{section_final_sample}.
\subsection{Preliminary cluster selection}
\label{clusterselection}
Several methods for identifying clusters of galaxies have been developed (e.g.,~\citealt{Abell58,Abell89,Zwicky61,Gunn86,Vikhlinin98,Kepner99,Gladders00,Piffaretti11,Planck2016}). However, each of them contains assumptions about cluster properties and is subject to selection effects. The earliest method used to identify clusters was the analysis of the optical images for the presence of over-density regions. Finding clusters with this method suffers from contamination by foreground and background galaxies that produce the false effect of over-density, which becomes more significant for high redshift. To help reduce this projection effect, another method one can use is the Red Sequence Method (RSM). This method is based on the fact that galaxy clusters contain a population of elliptical and lenticular galaxies that follow an empirical relationship between their color and magnitude and form the so-called red sequence~\citep{Gladders00}. The projection of random galaxies at different redshifts is not expected to form a clear red sequence. The RSM also requires multicolor observations. Spectroscopic redshift measurements help tremendously in establishing which galaxies are cluster members; though even, then the triple value problem can lead to erroneous associations.

\begin{figure}
\begin{center}
\includegraphics[scale=0.5]{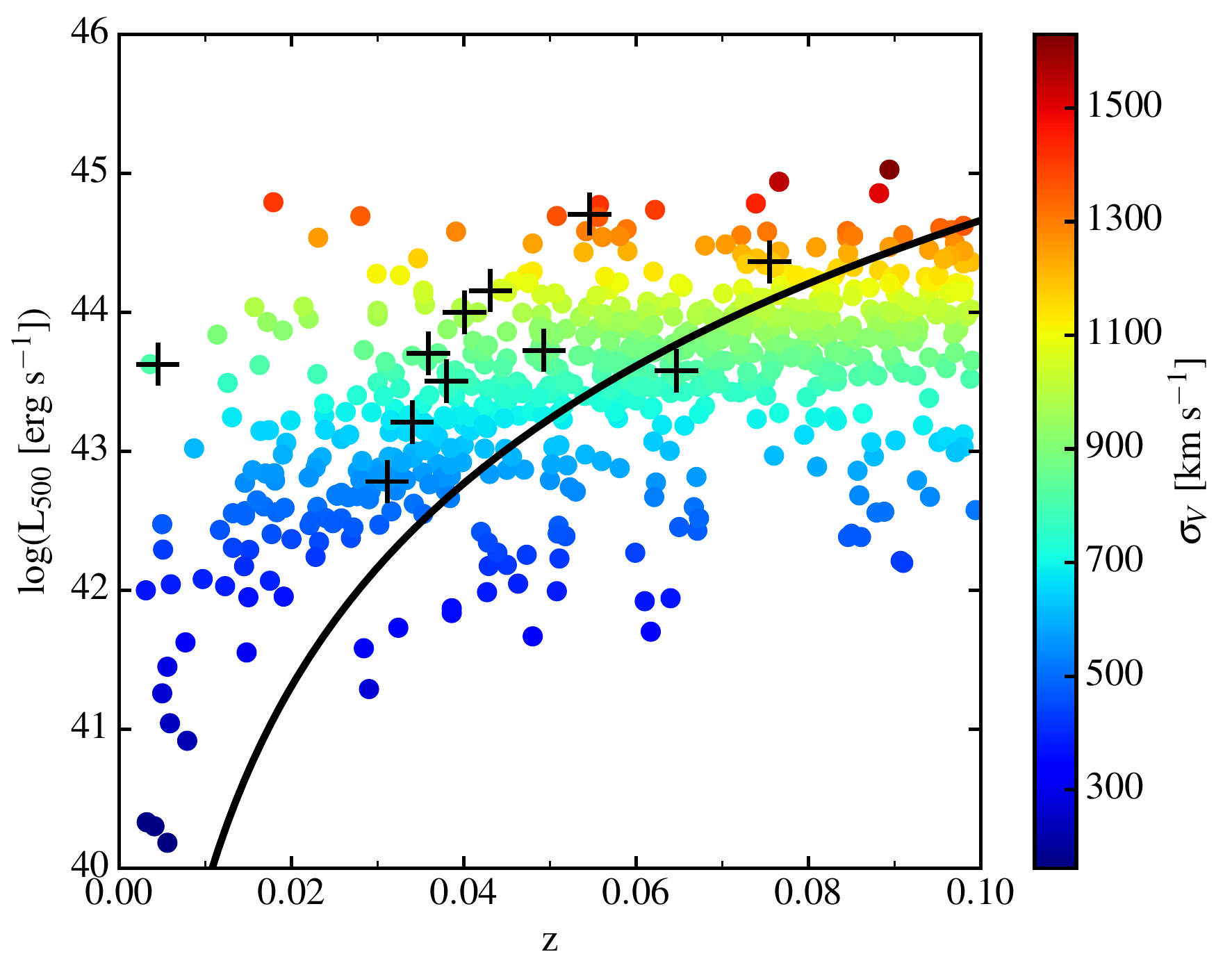}
\caption{The [0.1-2.4 keV] luminosities within $R_{500}$ of MCXC clusters~\citep{Piffaretti11} as a function of redshift, up to $z=0.1$. The colorbar shows the corresponding cluster velocity dispersion $\sigma_V$ calculated from~Eq.~\ref{sigmaV}. Black pluses are clusters from the current analysis. The black curve corresponds to the intrinsic dispersion of SNe~Ia on the Hubble diagram found for the JLA sample~\citep{Betoule14} projected onto cluster luminosities by combining the luminosity-mass and mass-velocity dispersion relations.}
\label{Lx}
\end{center}
\end{figure}

A third popular and effective method to detect galaxy clusters is to observe the diffuse X-ray emission radiated by the hot gas (10$^6$--10$^8$~K) in the centers of the clusters~\citep{Boldt66,Sarazin1988}. In virialized systems the thermal velocity of gas and the velocity of the galaxies in the cluster are determined by the same gravitational potential. As a result, clusters of galaxies where peculiar velocities are important appear as luminous X-ray emitters, with typical luminosities of $\cal{L}_{\mathrm{X}} \sim$ 10$^{43}$--10$^{45}$ erg s$^{-1}$. Such luminosities correspond to $\sigma_V \gtrsim 700$ km s$^{-1}$ (see Fig.~\ref{Lx}). The gas distribution can be rather compact and thus unresolved by X-ray surveys at intermediate and high redshifts. However, nearby clusters ($z<0.1$) will be well resolved, eliminating contamination from X-ray AGN or stars.

Finally, clusters of galaxies also cause distortions in the cosmic microwave background from the inverse Compton scattering of the CMB photons by the hot intra-cluster gas. In the fourth and final cluster identification method, this signature, known as the Sunyaev-Zel'dovich (SZ) effect, is used to identify clusters~\citep{Planck2016}. 

Using the \textsc{SIMBAD} database (\citealt{Wenger00}) we chose all the clusters projected within $\sim$2.5~Mpc around the SNe~Ia positions and with redshift differing from that of the supernova by less than 0.015. SN~Ia host redshifts were used to initially determine the distance. We did not consider objects classified as groups of galaxies (GrG), although there is no strong boundary between these and clusters, since groups of galaxies are characterized by smaller mass and therefore smaller velocity dispersion $\sim$300 km~s$^{-1}$~(see Fig.~5 in \citealt{2000ARA&A..38..289M}). The uncertainty introduced by such velocity is properly accounted for using the conventional method of assigning a fixed uncertainty to all SNe~Ia to account for peculiar velocities.

\subsection{Cluster redshift measurement}
\label{section_redshfit_measurement}
Some published cluster redshifts have been determined from a single or few galaxies. As we want to have a precise redshift correction, we can not simply replace the redshift of the host galaxy by the redshift of another galaxy. We therefore adopt the following methodology to improve cluster redshift estimates. 

To measure the redshift of the cluster it is necessary to know which galaxies in the cluster field are its members. Galaxy clusters considered in this paper are old enough ($z<0.1$) to exhibit virialized regions~\citep{2013ApJ...763...70W}. Therefore, to characterize the cluster radius we used the virial radius \rdeuxcent, corresponding to an average enclosed density equal to 200 times the critical density of the Universe at redshift~z:

\begin{equation}
\rdeuxcent \equiv R \rvert_{\overline{\rho}=200 \rho_c},
\end{equation}
\begin{equation}
\rho_c = \frac{3 H^2(z)}{8 \pi G},
\label{R200}
\end{equation}
where $H(z)$ is the Hubble parameter at redshift $z$ and $G$ is the Newtonian gravitational constant. 

According to the virial theorem, the velocity dispersion \sigv~inside a cluster is given as:
 
\begin{equation}
\sigv  \approx \sqrt{\frac{G \mdeuxcent}{\rdeuxcent}}.
\end{equation}
Using Eq.~\ref{R200} and $M_{200} = \frac{4}{3}\pi R_{200}^3 200\rho_{c}$ we find:

\begin{equation}
\sigv \approx 10\, \rdeuxcent\, H(z).
\label{sigmaV}
\end{equation}

The cluster redshift uncertainty \textbf{($z^{\,cl}_{err}$)} can be found from the cluster velocity dispersion:

\begin{equation}
z^{\,cl}_{err} = \frac{\sigv}{\sqrt{N_{gal}}}, 
\label{z_error}
\end{equation}
where $N_{gal}$ is a number of cluster members used for the calculation.

First, we took all the galaxies attributed to each cluster in literature sources and added the \textsc{SNfactory} host galaxy if it was not among them. Then, these data were combined with the DR13 release database of SDSS \citep{Eisenstein11,Dawson13,Smee13,DR13}. We selected all galaxies with spectroscopic redshifts located in a circle with the center corresponding to the cluster coordinates and projected inside the cluster's $R_{200}$ radius. A~$5\sigma_V$ redshift cut was adopted in the redshift direction (see~Eq.~\ref{sigmaV}). \\

The $R_{200}$ value was extracted from the literature when possible. For the clusters without published size measurements we estimated $R_{200}$ ourselves from the velocity distribution of galaxies around the cluster position following the procedure described in~\cite{Beers90} with an initial guess of $R_{200}=1.1 \ \text{Mpc}$. If the number of cluster members with spectroscopically determined redshifts was less than ten, the value of 1.1 Mpc was adopted as a virial radius. This value corresponds to the average $R_{200}$ of clusters in the MCXC, a meta-catalogue of X-ray  detected  clusters  of  galaxies (\citealt{Piffaretti11}); see Fig.~\ref{Lx}.

To estimate the redshift of a cluster we applied the so-called bi-weight technique~\citep{Beers90} on the remaining redshift distributions. Bi-weight determines the kinematic properties of galaxy clusters while being resistant to the presence of outliers and is robust for a broad range of underlying velocity distributions, even if they are non-Gaussian, using the median and an outlier rejection based on the median absolute deviation. Moreover, \citealt{Beers90} provide a formula for the cluster redshift uncertainty, but it can not be used for clusters with few members. Therefore, instead we use  Eq.~\ref{z_error}, which can be applied for all of our clusters.

\begin{figure*}
\begin{center}
\includegraphics[scale=0.6]{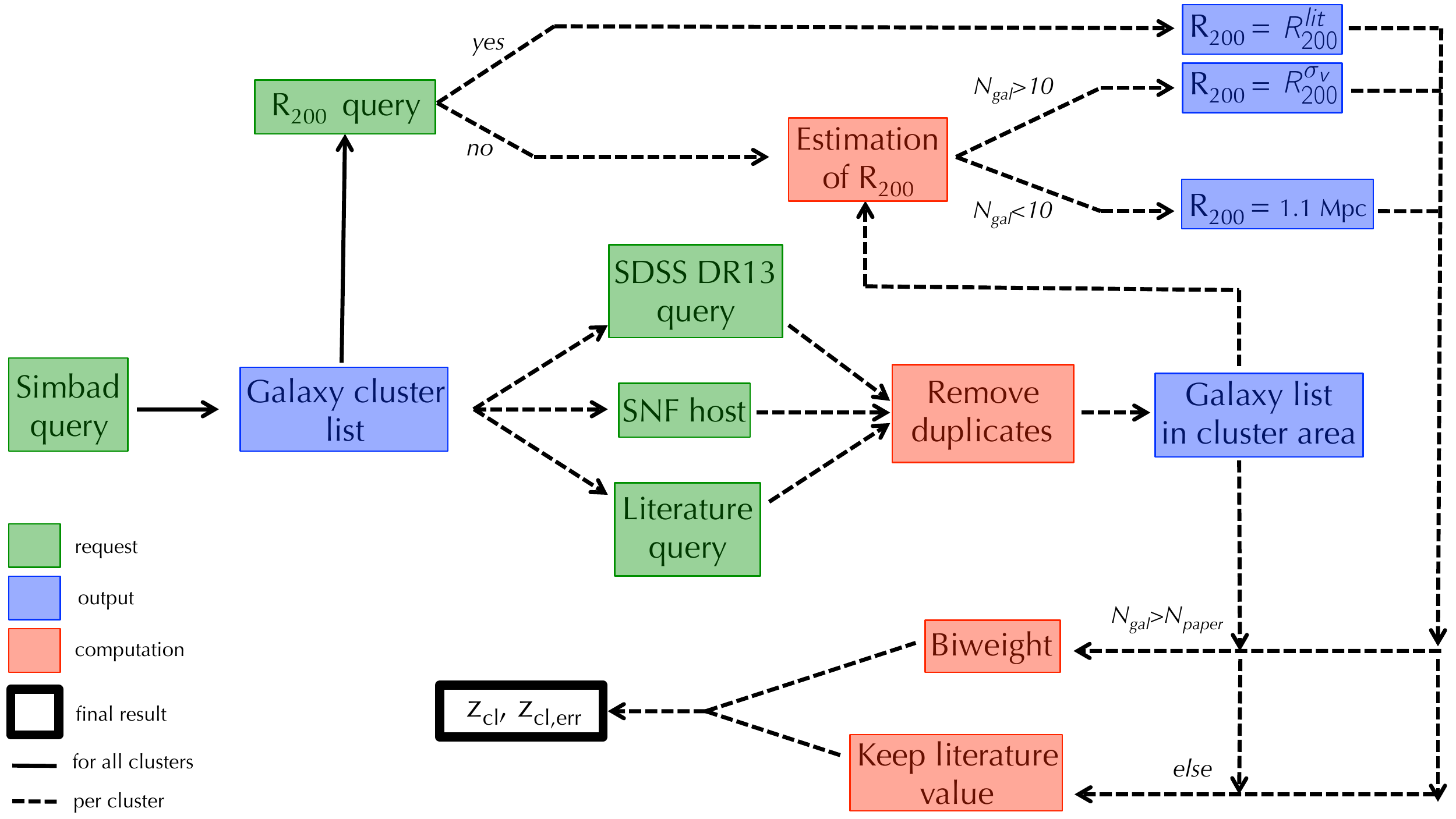}
\caption{Workflow for redshift calculation and other inputs for matching of SNe to galaxy clusters. In this scheme, $N_{gal}$ corresponds to the number of galaxies used to compute the redshift and $N_{paper}$ corresponds to the number of galaxies used to estimate the redshift in the literature.}
\label{redshift_computation}
\end{center}
\end{figure*}

For some of the clusters the literature provides only the final redshift and the number of galaxies, $N_{paper}$, that were used in the calculation, without publishing a list of cluster members. In those cases, if the number of members collected by us satisfies $N_{gal} < N_{paper}$ we adopted the redshift from literature. The detailed scheme of the cluster redshift calculation is presented in Fig.~\ref{redshift_computation}.

All the calculations described above are based on spectroscopical redshifts. Before performing the calculations of the cluster CMB redshift, all of the heliocentric redshifts of its members were first transformed to the CMB frame. The transformation to the CMB frame made use of the NASA/IPAC Extragalactic Database (\textsc{NED}).

\subsection{Final matching and confirmation}
\label{section_final_sample}
Once the redshifts and $R_{200}$ values were obtained for each cluster, we performed the final matching. A supernova is considered a cluster member if two conditions are satisfied:
\begin{itemize}
\item[$\bullet$] $r < R_{200}$, where $r$ is the projected distance between the SN and cluster center.
\item[$\bullet$] $|z^{\,host}-z^{\,cl}| < 3 \frac{\sigv}{c}$
\end{itemize}
The SNe~Ia that did not satisfy these criteria were removed from further consideration. 

Our final criteria are slightly different than those applied by \citealt{Xavier2013} (1.5 Mpc and $\sigma_V = 500$ km~s$^{-1}$) and \citealt{Dilday2010} (1 Mpc $h^{-1}$ and $\Delta z = 0.015$). They studied the properties and rate of supernovae in clusters and their choices were made to be consistent with previous cluster SN~Ia rate measurements. These values roughly characterize an ``average'' cluster and we were guided by the same thoughts when making the preliminary cluster selection (2.5 Mpc and $\Delta z =0.015$, see Sec.~\ref{clusterselection}). However, since clusters have different size and velocity dispersion, we determined or extracted from the literature the physical parameters of each cluster ($R_{200}$ and $\sigma_V$). This method provides an individual approach to each SN-cluster pair and allows association with a cluster to be defined with greater accuracy. 

\begin{figure}
\begin{center}
\includegraphics[scale=0.37]{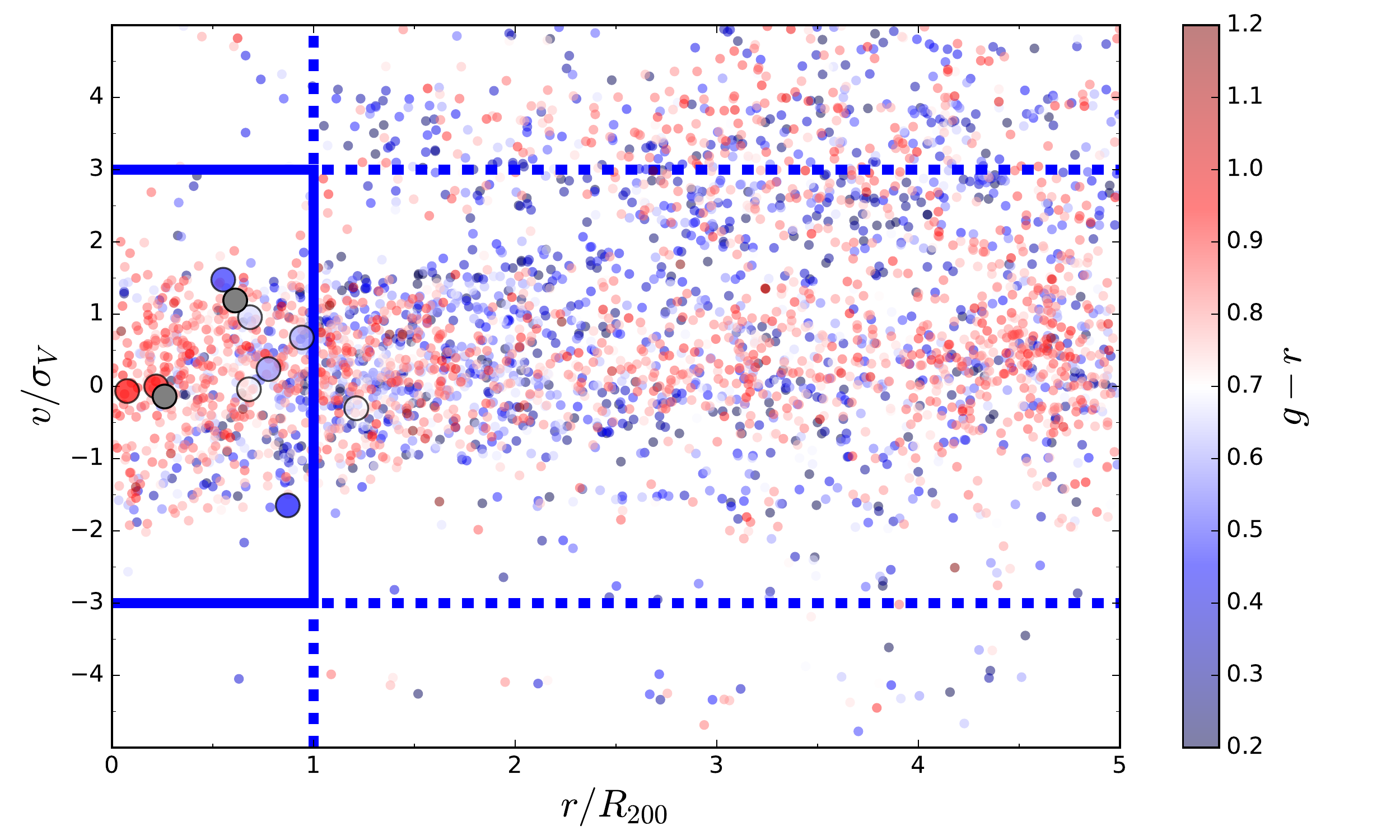}
\caption{Speed normalized by velocity dispersion within the ensemble cluster vs. the distance between galaxies and the ensemble cluster normalized by $R_{200}$. The small points are galaxies with spectroscopy from SDSS. The big points represent the positions of host galaxies of our SNe~Ia. The color bar shows the corresponding $g-r$ color, the points filled with grey do not have color measurements. The solid lines show the cuts we applied to associate SNe~Ia with clusters, and the dashed lines represent the prolongation of those cuts.}
\label{v_vs_distance}
\end{center}
\end{figure}

Following~\cite{1997ApJ...478..462C} and \cite{Rines2006} we constructed an ensemble cluster from all the clusters associated with SNe~Ia to smooth over the asymmetries in the individual clusters. We scaled the velocities by $\sigma_V$ and positions with the values of $R_{200}$ for each cluster to produce the Fig.~\ref{v_vs_distance}. This shows our selection boundaries and exhibits good separation of cluster galaxies from surrounding galaxies.

As it was mentioned in Sect.~\ref{clusterselection} there are several methods to identify a cluster. Initially we considered everything that is classified as a cluster by previous studies. However, some of these classifications can be false. For the remaining clusters we checked for the presence of X-ray emission, a red sequence or the SZ effect, as described below.

We used the public ROSAT All Sky Survey images within the energy band 0.1-2.4~keV, to look for extended X-ray counterparts\footnote{\url{http://www.xray.mpe.mpg.de/cgi-bin/rosat/rosat-survey}}. The expected [0.1-2.4 keV] luminosity within $R_{500}$ can be extracted from the luminosity-mass relation $h(z)^{-7/3}\left(\frac{L_{500}}{10^{44} erg s^{-1}}\right)= C\left(\frac{M_{500}}{3\times10^{14} M_\odot}\right)^\alpha$ with $\rm{log}(C)=0.274$ and $\alpha=1.64$ (see Table 1 in~\citealt{Arnaud2010}). The $L_{500}$ values for MCXC clusters~\citep{Piffaretti11} as a function of redshift are presented in Fig.~\ref{Lx}. Moreover, in Fig.~\ref{Lx} we have represented by a continuous black line the minimum value of $L_{500}$ which is required for the velocity dispersion of the cluster to cause a deviation from the Hubble diagram greater than the intrinsic dispersion in luminosity of SNe~Ia. It can be seen that all the clusters hosting SNe~Ia except one are above this threshold and it is therefore very likely that the Doppler effect induced by these clusters causes a dispersion in the Hubble diagram which is greater than the intrinsic dispersion in luminosity of SNe~Ia. Moreover, more than a half of the low redshift clusters are above this limit, indicating that the peculiar velocity correction has to be taken into account if a SN~Ia belongs to a cluster of galaxies and is observed at low redshift. 

To check for a linear red sequence feature, SDSS data were employed. From the SDSS \texttt{Galaxy} table we chose all the galaxies in the $R_{200}$ region around the cluster position. We extracted \texttt{model} magnitudes, as recommended by SDSS for measuring colors of extended objects. 

We checked for detections of the SZ effect using the Planck catalog of Sunyaev-Zel'dovich sources~\citep{2016A&A...594A..27P}. All the clusters in our sample with SZ sources also have X-ray emission, as expected for real clusters. 

Some of our supposed clusters do not show X-ray or SZ signatures of a cluster. As described in Sect.~\ref{clusterselection}, low-redshift clusters are expected to have X-ray emission. Therefore, only such candidates were kept for further analysis (see~Fig.~\ref{Lx}). 

Cases when the red sequence is clearly seen but for which there is no diffuse X-ray emission can be explained either by the superposition of nearby clusters or being a group embedded in a filament. For example, our study of the redshift distribution and sky projection around the proposed host cluster [WHL2012] J132045.4+211627 of SNF20070417-002 revealed that many of the redshifts used to determine $z^{\,cl}$ come from galaxies that are more spread out --- like a filament would be. We conclude that, consistent with the lack of X-rays, this is not a cluster.

Two other clusters, ZwCl 2259+0746 and A87, also require discussion. Within $2.3'$ of the center of ZwCl 2259+0746 there is a source of X-ray emission, 1RXS J230215.3+080159. However, the size of the emission region ($3'$) is very small in comparison with $R_{500}$ value for the cluster ($40'$). In addition, according to~\cite{2006A&A...449..425M} this emission belongs to a star. Therefore, we did not assign this X-ray source to ZwCl 2259+0746. Another case is A87, which belongs to the A85/87/89 complex of clusters of galaxies. According to~\cite{1998A&A...335...41D} the galaxy velocities in the A87 region show the existence of subgroups, which all have an X-ray counterpart, and seem to be falling onto A85 along a filament. Therefore, A87 is not really a cluster but a substructure of A85 that has a very prominent diffuse X-ray emission \citep{Piffaretti11}. We applied our redshift measurement technique to determine the CMB redshift of the virialized region of A85. Thus, we included A85/A87 in our final table for the peculiar velocity analysis.

The final list of  \textsc{SNfactory} SNe~Ia in confirmed clusters contains \NsnCl objects. The resulting association of SNe~Ia with host clusters is given in Table~\ref{cluster_list}. Column~1 is the SN name, Col.~2 contains a name of the identified host cluster of galaxies, and Col.~3 is the MCXC name. The MCXC coordinates of the host cluster center are given in Col.~4. Column~5 contains the projected separation, $D$, in Mpc between the SN position and the host cluster center. The $R_{200}$ value is in Col.~6 and the CMB supernova redshift is in Col.~7. The CMB redshift of the cluster and its uncertainty can be found in Cols.~8 and~9. The velocity dispersion of the cluster estimated from the $R_{200}$ value is shown in Col.~10. The number of galaxies that were used for cluster redshift calculation is in Col.~11. In Col.~12 we indicate the source of galaxy redshift information (lit. is an abbreviation for literature). In the last Col. we summarize all references for the cluster coordinates, $R_{200}$, and non-SN galaxy redshifts.

\begin{table*}
\begin{adjustbox}{max width=\textwidth} 
\begin{tabular}{llllp{1cm}p{1cm}lllp{1cm}rrr}
   \hline \hline
SN Name  & Host Cluster & MCXC Name &  Cluster Coordinates &  r (Mpc) & $R_{200}^{1}$ (Mpc)& $z^{\,host}_{c}$ & $z^{\,cl}_{c}$ & $z^{\,cl}_{er_{R200}}$ & $\sigma_{V_{R200}}$ (km/s) & $N_{gal}$ & Source & Ref.\\
  \hline  
SNF20051003-004/SN2005eu & RXJ0228.2+2811    &J0228.1+2811 & 02 28 09.6 +28 11 40                     & 0.24 & 0.92 & 0.0337 & 0.0340 & 0.0015 &644 & 2   & lit.      & 1,2,3\\
SNF20060609-002          & A2151a            &J1604.5+1743& 16 04 35.7 +17 43 28                     & 0.64 & 1.16 & 0.0399 & 0.0359 & 0.0002 &812 & 146 & SDSS+lit. & 1,4\\
SNF20061020-000          & A76               &J0040.0+0649 & 00 40 00.5 +06 49 05                     & 0.72 & 1.06 & 0.0379 & 0.0380 & 0.0008 &742 & 9   & SDSS+lit. & 1,5\\
SNF20061111-002          & RXC J2306.8-1324  &J2306.8$-$1324   & 23 06 51.7 $-$13 24 59                   & 0.66 & 1.08 & 0.0677 & 0.0647 & 0.0018 &756 & 2   & lit.      & 1,6\\
SNF20080612-003          & RXC J1615.5+1927  &J1615.5+1927& 16 15 34.7 +19 27 36                     & 0.52 & 0.76 & 0.0328 & 0.0311 & 0.0004 &532 & 19  & SDSS      & 1 \\
SNF20080623-001          & ZwCl8338          &J1811.0+4954  & 18 11 00.1 +49 54 40                     & 1.02 & 1.17 & 0.0448 & 0.0493 & 0.0005 &819 & 36  & lit.      & 1,4\\
SNF20080731-000          & ZwCl 1742+3306    &J1744.2+3259  & 17 44 15.0 +32 59 23                     & 0.34 & 1.55 & 0.0755 & 0.0755 & 0.0026 &1085& 2   & lit.      & 1,7\\
PTF09foz                 & A87/A85           &J0041.8$-$0918& 00 41 50.1 $-$09 18 07                 & 2.23 & 1.84 & 0.0533 & 0.0546 & 0.0004 &1288& 148 & SDSS      & 1,8\\
SN2006X                  & Virgo             &J1230.7+1220 & 12 30 47.3 +12 20 13                     & 1.19 & 1.14 & 0.0063 & 0.0045 & 0.0001 &798 & 607 & SDSS+lit. & 1,9,10     \\    
SN2007nq                 & A119              &J0056.3$-$0112& 00 56 18.3 $-$01 13 00                 & 1.11 & 1.43 & 0.0439 & 0.0431 & 0.0003 &1001& 153 & SDSS+lit. & 1,4,11\\
SN2009hi                 & A2589             &J2323.8+1648& 23 23 53.5 +16 48 32                     & 0.10 & 1.33 & 0.0399 & 0.0401 & 0.0004 &931 & 54  & SDSS+lit. & 1,4,12\\
  \hline
  \end{tabular}
      \end{adjustbox}
  \caption{\label{tab:robots}The association of the \textsc{SNfactory} SNe~Ia with host clusters. $^1$The value was calculated from $R_{500}$ with equation $R_{200}=1.52 R_{500}$~\citep{Reiprich2002,Piffaretti11}. References. (1) \citealt{Piffaretti11}; (2) \citealt{Wegner93}; (3) \citealt{2005IAUC.8611....1L}; (4) \citealt{Smith04}; (5) \citealt{Hudson01}; (6) \citealt{Cruddace02}; (7) \citealt{Ulrich76}; (8) \citealt{2009ATel.2255....1N}; (9) \citealt{2006IAUC.8667....1P}; (10) \citealt{2014ApJS..215...22K}; (11) \citealt{2007CBET.1106....1Q}; (12) \citealt{2009CBET.1872....1N}.}
  \label{cluster_list}
\end{table*}

\section{Impact on the Hubble diagram}
\label{impactHD}
Since we have a list of \NsnCl~SNe~Ia that belong to clusters, we can apply peculiar velocity corrections and study how they affect the Hubble residuals. The following methodology is implemented. 

The theoretical distance modulus is $\mu^{\,th} = 5\log_{10} d_{\rm L} - 5$, where $d_{\rm L}$ is the true luminosity distance in parsecs:

\begin{equation}
 d_{\rm L} = \frac{c}{H_0}(1+z_{\rm h})\int_{0}^{z_c} \frac{dz'_c}{\sqrt{\Omega_\Lambda + \Omega_m(1 + z'_c)^3}},
\end{equation}
where $z_{\rm h}$ is the heliocentric redshift, which takes into account the fact that the observed flux is affected not only by the cosmological redshift but by the Doppler effect as well.

We assign the cosmological redshift $z_{c}$ to be:

\begin{equation}
\label{second_correction}
z_{c} = \left\{
  \begin{array}{ll}
    z^{\,cl}_{c} & \text{if inside a galaxy cluster,}  \\
     & \\
    z^{\,host}_{c} & \text{otherwise.} \\
  \end{array}
\right.
\end{equation}

\noindent The uncertainty on $z_{c}$ (both SN~Ia and host cluster) is propagated into the magnitude error ${\sigma_i^{\,tot}}^2$ as:

\begin{equation}
{\sigma_i^{\,tot}}^2 = \sigma_{LC_i}^{\,2} + \ \sigma_z^{\,2} + \ \sigma_{int}^{\,2}
\end{equation}
where $\sigma_{LC_i}$ is the propagation of uncertainty from light curve parameters to an apparent magnitude of SN~Ia in the $B$-band $\mbstar$, $\sigma_{int}$ is the unknown intrinsic dispersion of SN~Ia. $\sigma_z$ is the uncertainty on redshift measurement and peculiar velocity correction (see~Eq.~\ref{erreur_vitesse_prope}), which is assigned as: 

\begin{equation}
\label{third_correction}
\sigma_z=\left\{
  \begin{array}{ll}
    \frac{5 \ \sqrt{z^{\,cl \ 2}_{err}}} {z^{\,cl} \ln(10)}& \text{if inside a galaxy cluster,}  \\
     & \\
    \frac{5 \ \sqrt{z^{\,host \ 2}_{err}+0.001^2}} {z^{\,host} \ln(10)} & \text{otherwise.} \\
  \end{array}
\right.
\end{equation}
The 0.001 value corresponds to the 300~km~s$^{-1}$~that is added to the redshift error of SNe~Ia outside the clusters in order to take into account the unknown galaxy peculiar velocities, as in a classical cosmological analysis. For cases where a SN~Ia belongs to a galaxy cluster, we assume that the redshift error contains only the error due to the redshift measurement of a cluster.

By fitting the Hubble diagram using only SNe~Ia outside the galaxy clusters\footnote{Taking into account all the SNe~Ia does not affect the Hubble diagram fitting because the number of SNe~Ia inside galaxy clusters is small.}, we obtained SN~Ia SALT2 nuisance parameters: $\alpha$ and $\beta$, the classical standardization parameters for light curve width and color respectively; the absolute magnitude $M_B$, and the intrinsic dispersion. These nuisance parameters remained fixed during our analysis. Once the nuisance parameters were estimated, we computed the difference between observed and theoretical distance modulus (Hubble residuals). In order to study the impact of peculiar velocity correction, we compute the Hubble residuals for the SNe~Ia in clusters before and after correction. We used the weighted root mean square  (\wrms) as defined in \citealt{Blondin11} to measure the impact of this correction. We used the same intrinsic dispersion established during the fitting ($\sigma_{int}=0.10$~mag) to calculate all \wrms. SN2006X is not taken into account during the computation of the \wrms due to the fact that it does not belong to the set of ``normal'' SNe~Ia. However, SN2006X is included in the statistical tests described below (for details see Sect.~\ref{SN2006X_discusion}).

\begin{figure*}
\begin{center}
\includegraphics[scale=0.4]{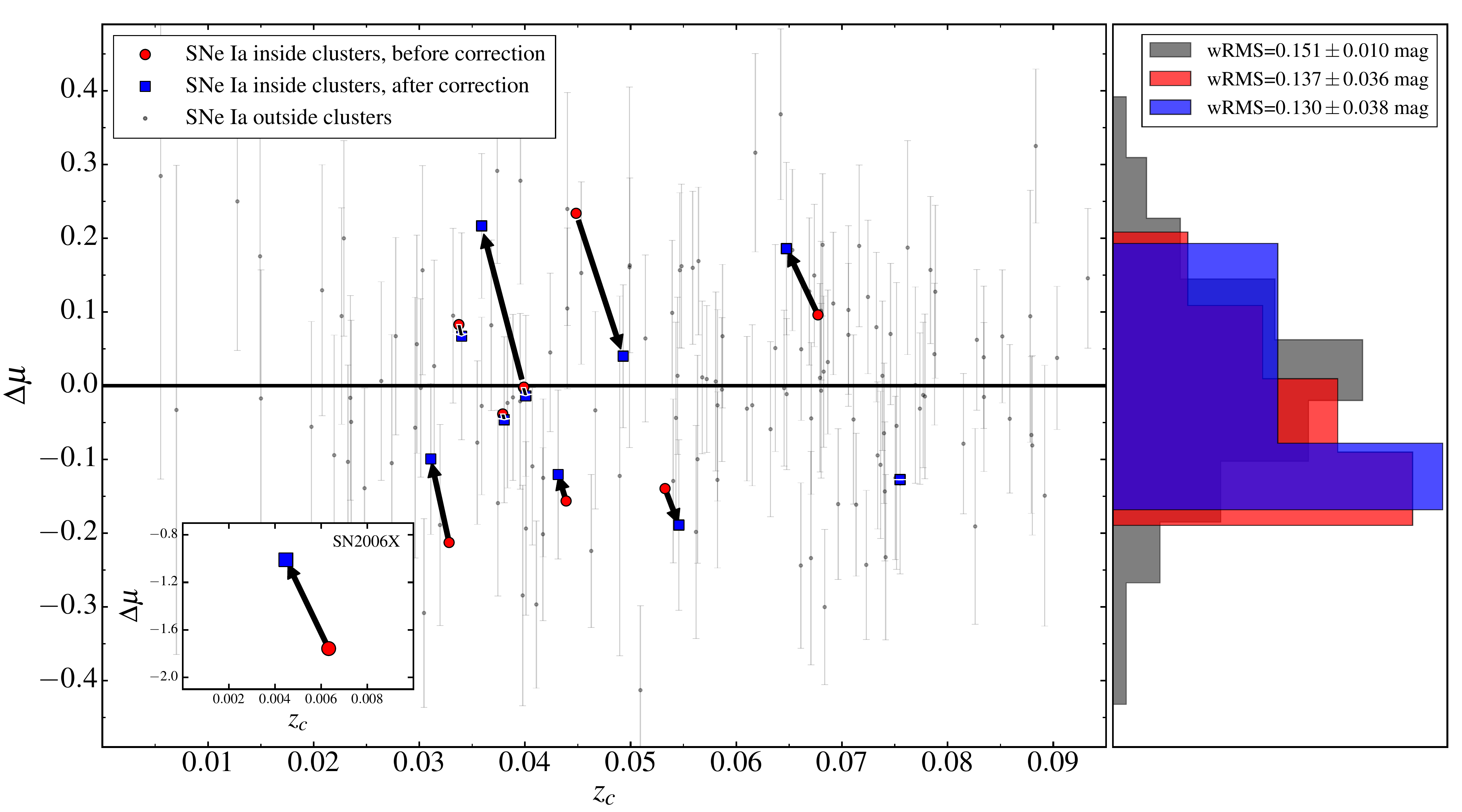}
\caption{Hubble diagram residuals. For cluster members red circles (blue squares) and histograms correspond to residuals for \sne~in galaxy clusters before (after) correction for peculiar velocities of the hosts inside their clusters. The black histogram corresponds to all SNe~Ia after correction. SN2006X is presented in the inset plot separately from the others due to its very large offset.}
\label{hubble_corrigeR200_zoom}
\end{center}
\end{figure*}

\begin{figure}
\begin{center}
\includegraphics[scale=0.45]{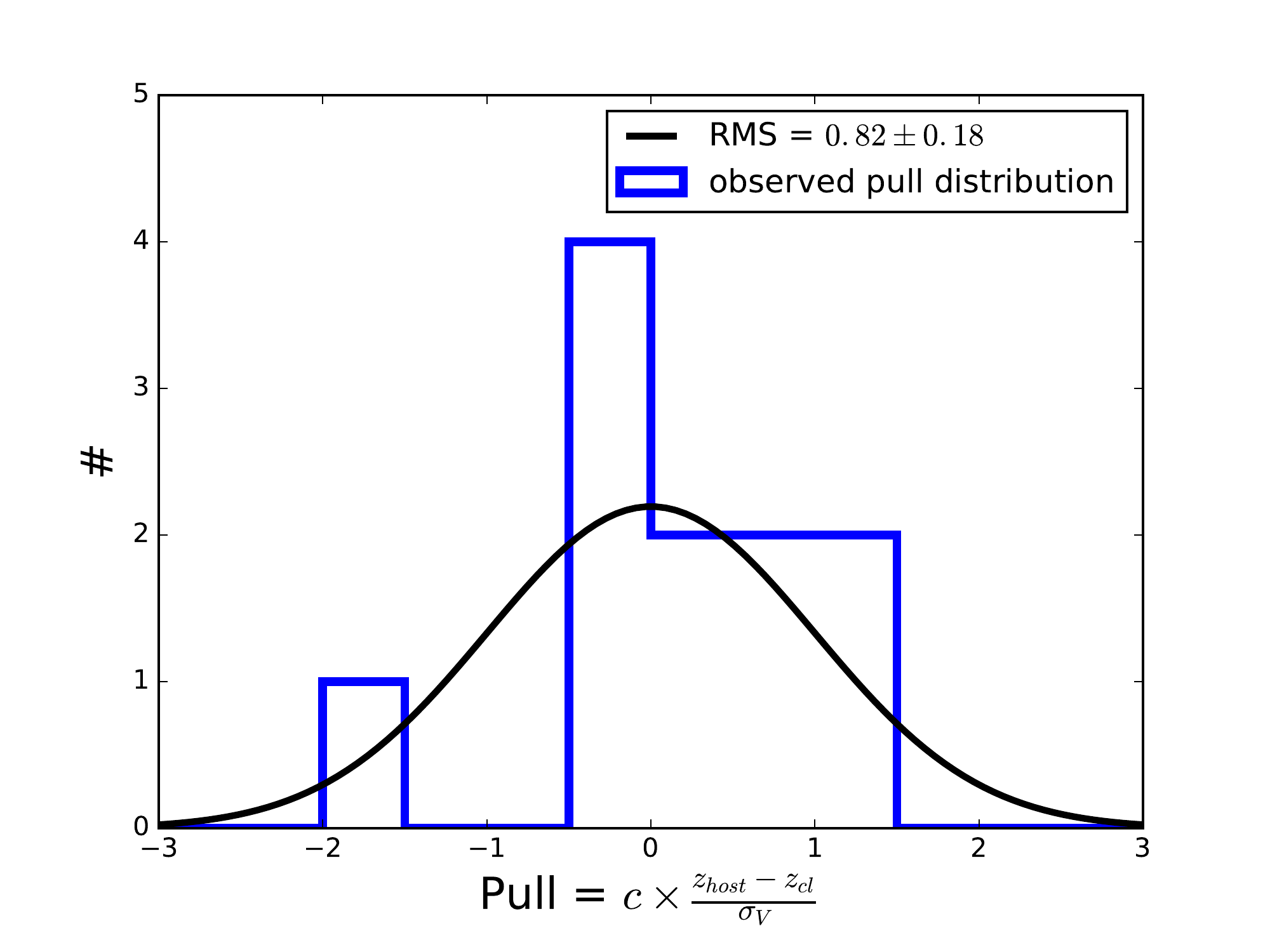}
\caption{Velocity pull distribution (in blue) in comparison with a Gaussian distribution with the observed standard distribution of the velocity pull. This is compatible with the expected standard deviation of unity.}
\label{velocity_pull}
\end{center}
\end{figure}

\begin{figure}
\begin{center}
\includegraphics[scale=0.45]{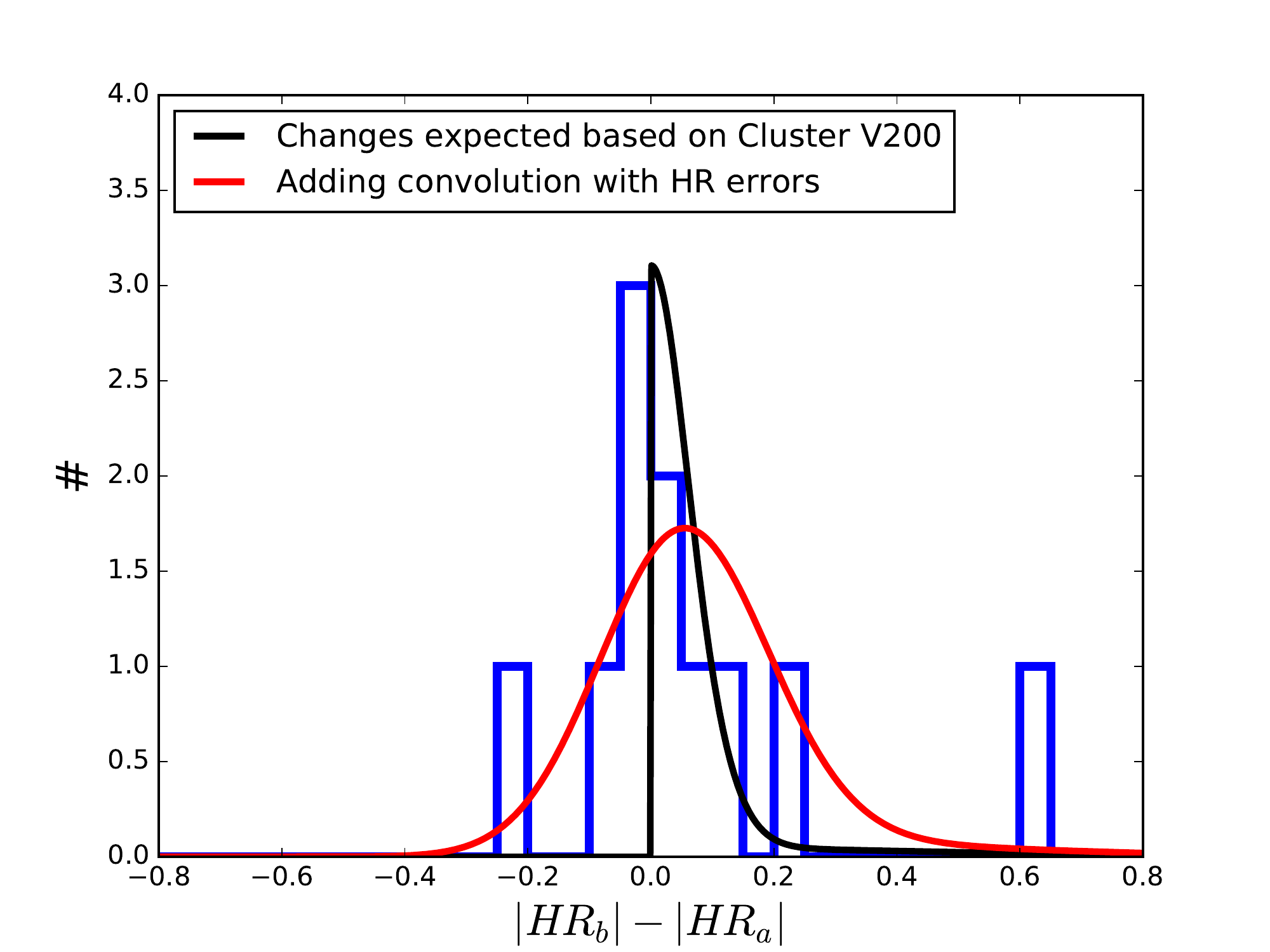}
\caption{Distribution of the difference in absolute HR after ($HR_a$) and before ($HR_b$) peculiar velocity correction (in blue). The black line represents the expected distribution of the difference in HR, and the red curve is the expected change convolved with error distribution. The results are compatible with the observed distribution given the Poisson uncertainties of each histogram bin.}
\label{expected_HR_distribution}
\end{center}
\end{figure}

The dispersion of these \NsnCl SNe~Ia around the Hubble diagram decreases significantly when the peculiar velocities of their hosts inside the clusters are taken into account (\wrms=\wrmsClusDoppler~mag). When using the redshift of the host instead of the redshift of the cluster, the dispersion of these \NsnCl SNe~Ia is \wrms=\wrmsClusNoDoppler mag (see~Fig.~\ref{hubble_corrigeR200_zoom}). In order to compute the significance of this improvement, the Pearson correlation coefficient and its significance between HR before the correction and $5\log_{10}\left(z^{\,cl}/z^{\,host}\right)$ are computed. The Pearson correlation coefficient is \pearsonCluster, and its significance is \pearsonsigCluster, which is significant. In order to crosscheck this significance, we did a Monte-Carlo simulations. For each simulation, we took the difference $z^{\,cl}-z^{\,host}$ for the \NsnCl SNe~Ia in clusters and then randomly applied these corrections to the same \NsnCl SNe~Ia. For each simulation, we examine how often we get a \wrms less than or equal to the observed \wrms after the fake random peculiar velocity correction. On average the \wrms is higher and the probability to have the same or lower dispersion in \wrms is \pvalueCluster, which is in agreement with Pearson correlation significance. \\
\indent Even though the p-value is low, we still need to clarify why the decrease in wRMS is not higher. In order to examine whether the corrections are consistent with what it is expected, we compute the distribution of the pull of peculiar velocities and the expected distribution of HRs of our correction. These two distributions are shown respectively in Fig.~\ref{velocity_pull} and Fig.~\ref{expected_HR_distribution}. For the pull distribution shown in Fig.~\ref{velocity_pull}, which is defined as the distribution of difference between the host galaxy redshift and the host galaxy clusters redshift, divided by the peculiar velocity dispersion within the cluster, we should expect to get a centered normal distribution with a standard deviation of unity. The standard deviation of the pull is $0.82 \pm 0.18$ which is consistent with the expected unity distribution of the pulls. In addition, we showed in Fig. \ref{expected_HR_distribution} the expected distribution of the correction, the expected distribution of the correction convolved with uncertainties on HR, and the observed distribution of the correction. It is seen that the observed distribution of the corrections and the predicted distribution of the corrections are consistent.\\ 
\indent To resume, the Pearson correlation coefficient and its significance, the distribution of the pull, and the comparison between the expected correction and the observed correction show that our correction is consistent with expectations given the cluster velocity dispersions and the uncertainty in SN~Ia luminosity distance.

In addition, the \wrms we found for SNe~Ia inside the clusters before correction, \mbox{\wrmsClusNoDoppler~mag}, is also smaller than the \wrms for the SNe~Ia in the field (\wrms = \wrmsOutClusDoppler mag). This is consistent with a statistical fluctuations, but could be due to a lower intrinsic luminosity dispersion for SNe~Ia inside galaxy clusters. This possibility will be explored in the Sect.~\ref{Luminosity_dispersion_GC}.
 
\section{Discussion}
\label{discussion}

\subsection{SN2006X}
\label{SN2006X_discusion}

Throughout the analysis, we treated SN2006X in a special way because this SN~Ia is highly reddened SN, i.e. it is associated with dusty local environment~\citep{Patat2007}. This is a SN~Ia affecting the interstellar medium and which exhibits very high ejecta velocities and a light echo \citep{Patat09}. These special features put it very far off the Hubble diagram, and makes this SN unsuitable for cosmological analysis. However, it was included in the analysis here because  we are interested in the impact of peculiar velocities within galaxy clusters, not cosmology alone, and it passes the light curve quality criteria defined in~\cite{Guy10}. While SN2006X can bias the dispersion, only the difference between the residuals before correction for peculiar velocity and after correction for peculiar velocity is taken into account in the computation of the significance of the signal. This correction for SN2006X is around $\sim$550 km~s$^{-1}$ in velocity, and has a huge impact on magnitude at nearby redshift. In this case the $\sim$0.7 magnitude correction improves the dispersion on the Hubble diagram. Indeed, the original Hubble residual was measured as $\sim-1.7$ mag when using the host galaxy redshift instead of the galaxy cluster's redshift whereas the Hubble residual is $\sim-1.0$ mag. This correction is <50\% of the original offset, and smaller than the corrected residual from stretch and color only.

Considering the importance of the correction for SN2006X and the fact that this object is peculiar, it makes sense to calculate the significance of the peculiar velocity correction when SN2006X is not taken into account. Without SN2006X, the Pearson correlation coefficient decreases substantially to \pearsonClusterwithoutSNX, with a signifcance of \pearsonsigClusterwithoutSNX. Moreover, by re-doing the same Monte-Carlo simulation as in Sect.~\ref{impactHD} for the remaining cluster SNe~Ia, the p-value changes from \pvalueCluster to \pvalueClusterwithoutSNX,  which is in agreement with Pearson correlation significance. Thus, removing an object where the correction is large decreases the significance of the correction, especially given the small sample size. 

\subsection{Physical properties of SNe~Ia and their hosts in galaxy clusters}
\label{Luminosity_dispersion_GC}
In Sect.~\ref{impactHD} it was shown that the \wrms around the Hubble diagram for the SNe~Ia in clusters is less than for SNe~Ia in the field, which suggests that SNe~Ia in clusters might represent a more ``standard'' subclass of SNe~Ia (see Fig.~\ref{hubble_corrigeR200_zoom}). In order to compute the significance of this lower dispersion we perform \pvalueNsim Monte-Carlo simulations. For each simulation, we randomly select \NsnCl SNe~Ia in our sample and compute the \wrms.  For all the simulations we compute how often the dispersion is lower than the dispersion of \wrms=\wrmsClusDoppler mag observed inside clusters after the peculiar velocity correction. In this case, the probability to have such a low dispersion is \pvalueDispCluster, which is not significant. 

\begin{figure}
\begin{center}
\includegraphics[scale=0.45]{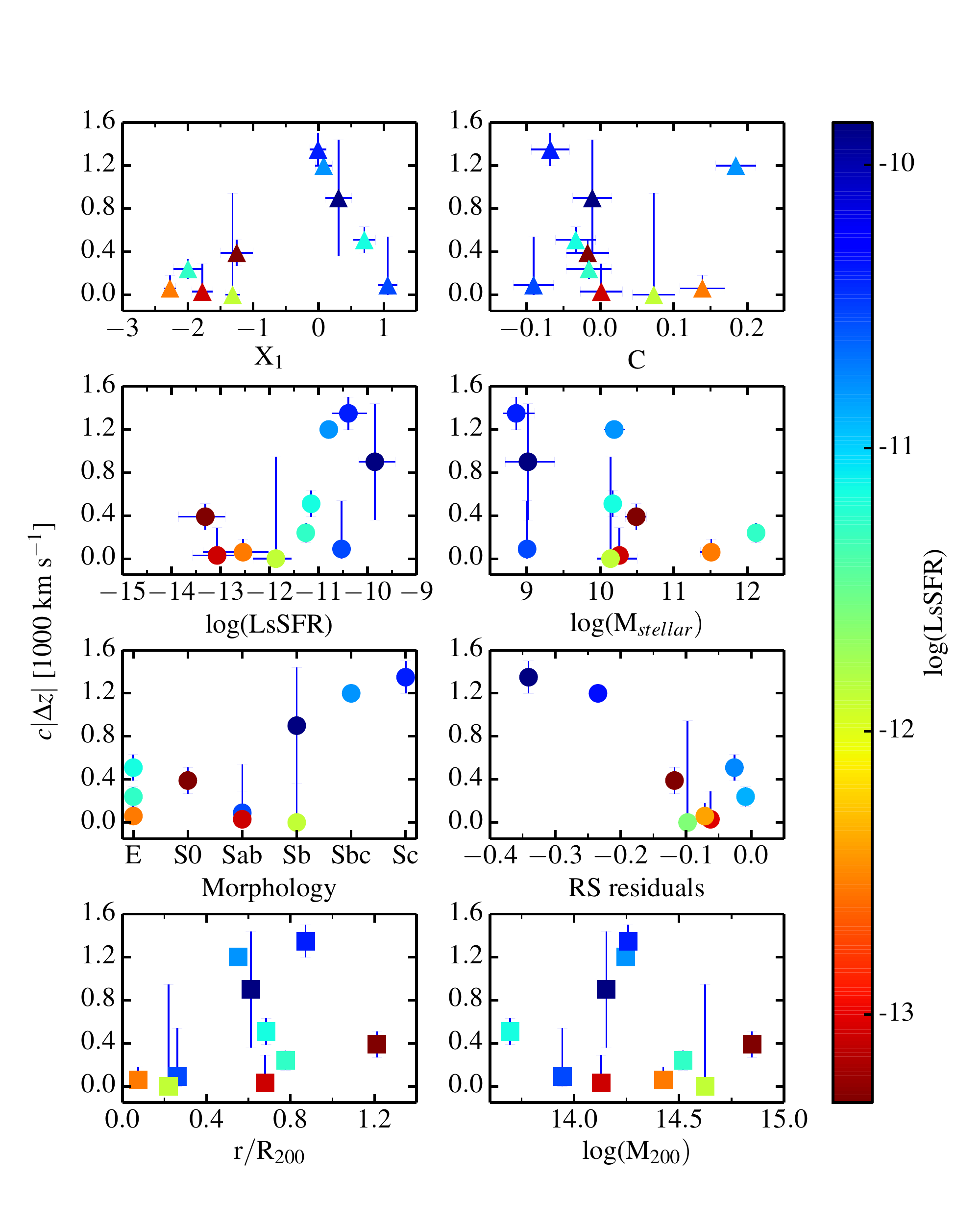}
\caption{Peculiar velocity correction, $c|\Delta z|$, for SNe~Ia that belong to clusters, as a function of supernova parameters ($X_1$, $C$; triangles), host properties (local specific SFR (yr$^{-1}$kpc$^{-2}$), stellar mass ($M_\odot$), morphological type, RS residuals [$(g-r) - (g-r)_{RS}$]; circles,  \citealt{2013ApJ...770..107C,Brown2014,Rigault18}), relative SN position inside the cluster and cluster mass~$M_{200}$ ($M_\odot$); squares. The colorbar shows the corresponding local specific SFR.} 
\label{fig:M200_vs_cdeltaz}
\end{center}
\end{figure}

Despite the low significance of their smaller dispersion, we could expect some difference between SNe~Ia in clusters and outside them because the properties of the galaxies inside the clusters are known to be different from those in the field. While in the field all morphological types of the galaxies are observed, the central parts of the clusters usually contain a large percentage of elliptical galaxies. The oldest stars, with an ages comparable to that of the Universe, lie in elliptical/lenticular galaxies. Moreover, dust is often absent in these regions. As shown by previous studies, narrow light curve SNe~Ia are preferentially hosted by galaxies with little or no ongoing star formation, and usually occur in more massive galaxies~\citep{1995AJ....109....1H,1996AJ....112.2398H,1999AJ....117..707R,2000AJ....120.1479H,2003MNRAS.340.1057S,2006ApJ...648..868S,2010MNRAS.406..782S,2009ApJ...707.1449N,2012ApJ...755...61S,2013MNRAS.435.1680J,2016arXiv161204417H,Henne2017}. Indeed, if we examine Fig.~\ref{fig:M200_vs_cdeltaz}, we see that \NsnCl SNe~Ia found in clusters are consistent with those studies, SNe~Ia with higher $X_1$ belong to the hosts with higher local sSFR and smaller $M_{stellar}$. The properties of 48 SNe~Ia in clusters vs. 1015 SNe~Ia in the field were studied in \cite{Xavier2013}, who found the following mean values for SN LC parameters: $\overline X_1 = 0.14\pm0.04$ (field), $-0.40\pm0.20$ (clusters) and $\overline C = -0.011\pm0.004$ (field), $-0.03\pm0.02$ (clusters). For comparison our means are $\overline X_1 = -0.01\pm0.09$ (field), $-0.65\pm0.36$/$-0.56\pm0.34$ (clusters without/with SN2006X) and $\overline C = 0.01\pm0.01$ (field), $0.02\pm0.03$/$0.13\pm0.11$ (clusters without/with SN2006X). The correlation between HRs for \NsnCl SNe in clusters and their host galaxy's mass is the same as shown in Fig. 15 of~\cite{Xavier2013}. 

\begin{figure*}
\begin{center}
\includegraphics[scale=0.6]{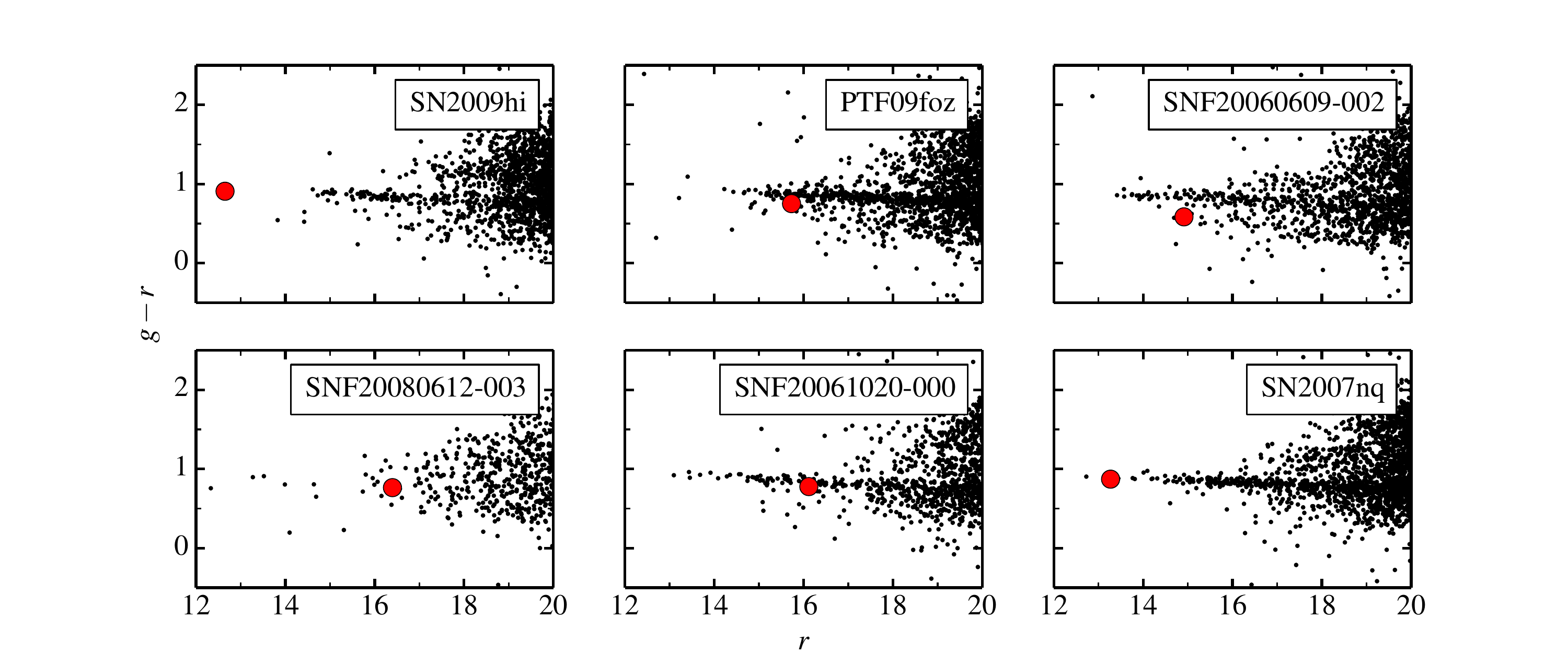}
\caption{Color-magnitude diagram ($g-r$ vs. $r$) plotted for the clusters from Table~\ref{cluster_list}  for the six clusters for which SDSS galaxy redshifts and colors are available \citep{Eisenstein11,Dawson13,Smee13,DR13}. Red points show the positions of supernova hosts, most of which are located near the red sequence.}
\label{RS}
\end{center}
\end{figure*}

We also performed morphological classification of the hosts (see Table~\ref{host_list}) based on the information provided by \textsc{SIMBAD} and \textsc{HyperLeda} databases~\citep{Wenger00,Makarov2014} and images from~\cite{2013ApJ...770..107C}. The host of SNF20080612-003 is classified as elliptical by \textsc{HyperLeda} and as spiral by~\cite{Sternberg11}. However, the classification by~\cite{Sternberg11} is based on images from Digital Sky Survey. This host looks elliptical without any sign of spiral arms on the SDSS image. Therefore, we assigned this galaxy to elliptical as in \textsc{HyperLeda}. We found that four SNe belong to elliptical/lenticular galaxies while the other seven are located in spirals. All of the early-type (elliptical and lenticular) galaxies fall on the red sequence for their clusters (see the color-magnitude diagrams ($g-r$ vs. $r$) for the clusters within the SDSS footprint in~Fig.~\ref{RS}). For the most part the spiral hosts are very close to the red sequence as well, i.e. these galaxies are characterized by redder colors. 

\begin{table*}
\centering
\begin{tabular}{lllrr}
   \hline \hline
SN Name & Host Name & Host Type & log(LsSFR) & log($M_{stellar}$)\\
\hline
SNF20051003-004 &    NSFJ022743.32+281037.6 &          Sab & $-10.53$ &  9.01 \\ 
SNF20060609-002 &             MCG+03-41-072 &          Sbc & $-10.79$ & 10.19 \\ 
SNF20061020-000 &    2MASXJ00410521+0647439 &          Sab & $-13.07$ & 10.26 \\ 
SNF20061111-002 &                         ... &           Sb &  $-9.85$ &  9.02 \\ 
SNF20080612-003 &    2MASXJ16152860+1913344 &            E & $-11.15$ & 10.17 \\ 
SNF20080623-001 &  WINGSJ181139.70+501057.1 &           Sc & $-10.39$ &  8.86 \\ 
SNF20080731-000 &                         ... &           Sb & $-11.87$ & 10.14 \\ 
PTF09foz        &    2MASXJ00421192-0952551 &           S0 & $-13.31$ & 10.49 \\ 
SN2006X         &                  NGC 4321 &           Sbc&      \textemdash &   \textemdash \\
SN2007nq        &                   UGC 595 &            E & $-11.26$ & 12.12 \\ 
SN2009hi        &                  NGC 7647 &            E & $-12.54$ & 11.51 \\ 
  \hline
\end{tabular}
  \caption{The properties of host galaxies of SNe~Ia belonging to galaxy clusters. LsSFR  (yr$^{-1}$kpc$^{-2}$) --- local specific star formation rate (star formation rate per unit galaxy stellar mass; \citealt{Rigault18}), $M_{stellar}$ ($M_\odot$) is the host galaxy stellar mass~\citep{2013ApJ...770..107C}.}
  \label{host_list}
\end{table*}

\begin{figure}
\begin{center}
\includegraphics[scale=0.45]{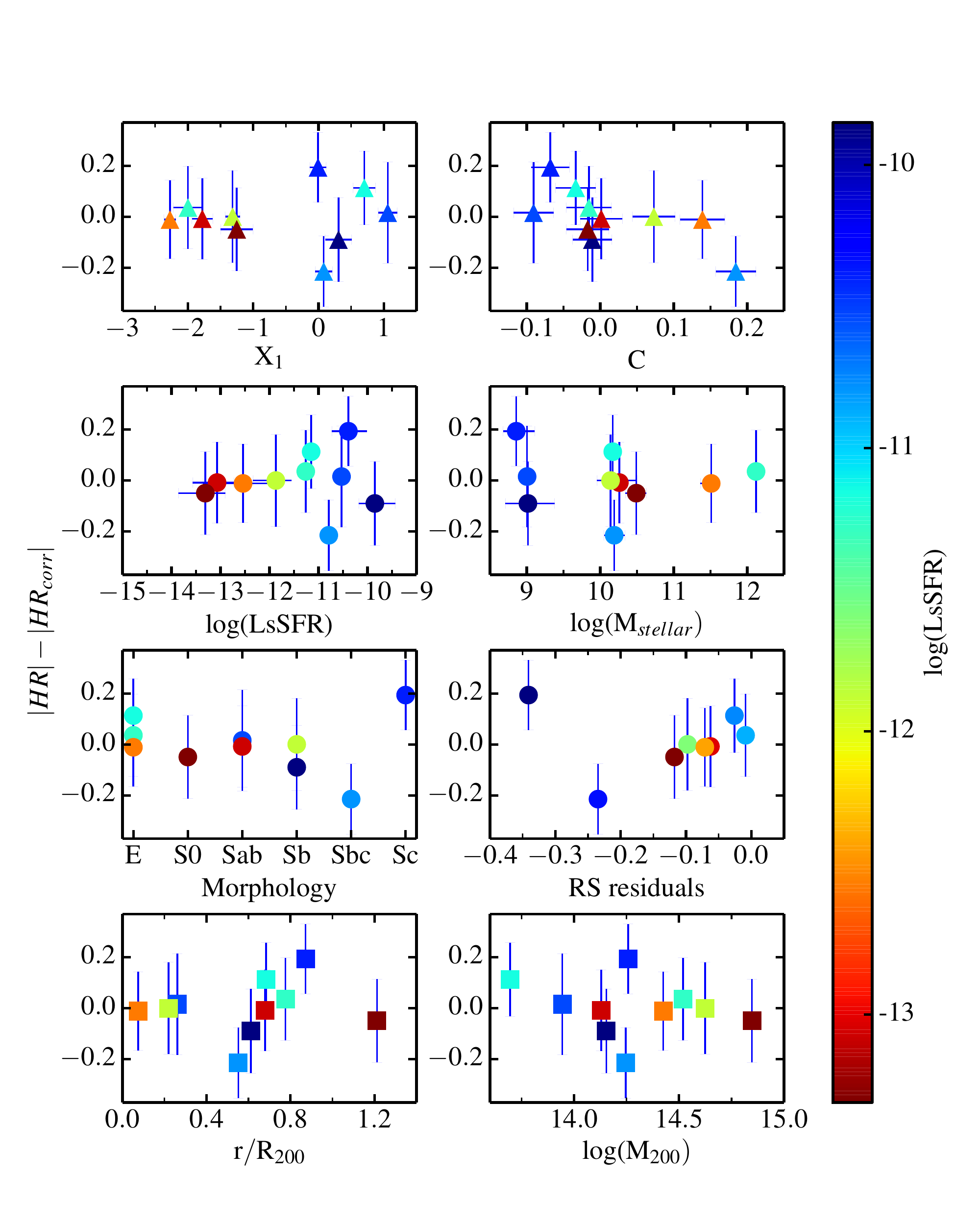}
\caption{Absolute change in Hubble residuals due to peculiar velocity correction for SNe~Ia that belong to clusters, as a function of supernova parameters ($X_1$, $C$; triangles), host properties (local specific SFR (yr$^{-1}$kpc$^{-2}$), stellar mass ($M_\odot$), morphological type, RS residuals [$(g-r) - (g-r)_{RS}$]; circles, \citealt{2013ApJ...770..107C,Brown2014,Rigault18}), relative SN position inside the cluster and cluster mass~$M_{200}$ ($M_\odot$); squares. The colorbar shows the corresponding local specific SFR.} 
\label{fig:M200_vs_residuals}
\end{center}
\end{figure}

In Figs.~\ref{fig:M200_vs_cdeltaz} and ~\ref{fig:M200_vs_residuals} we also show how the peculiar velocity correction $c|\Delta z|$ and the absolute change in Hubble residuals due to peculiar velocity correction depend on the supernova parameters $X_1$ and $C$, host properties such as local sSFR, stellar mass, morphological type, the difference between $(g-r)$ of the host and corresponding $(g-r)$ of the red sequence (RS residuals), relative SN position inside the cluster and cluster mass~$M_{200}$ (\citealt{2013ApJ...770..107C,Brown2014, Rigault18}). The $c|\Delta z|$ plot shows that most of the SNe whose redshifts are significantly changed have $X_1 \simeq 0$ and are hosted by blue spiral galaxies, having high local sSFR, smaller $M_{stellar}$, $r/R200\simeq0.7$ (see Fig.~\ref{fig:M200_vs_cdeltaz}). This is consistent with the distribution of galaxies in clusters such that the massive/elliptical/passive galaxies are located in the center but outer region contains spiral galaxies as well and with velocity profiles inside the clusters (see Fig.~\ref{v_vs_distance}, \citealt{1997ApJ...478..462C}, their Fig.~1, and \citealt{Rines2006}, their Fig.~15).

The small size of our sample does not allow us to perform cosmological fits separately for the SNe~Ia inside and outside galaxy clusters or to perform more detailed study of the behavior of the supernova light curve parameters in both subsamples. Once samples of SNe~Ia in clusters become much larger, it will be interesting to perform such analysis again, especially to find out whether variation of the light curves parameters, and luminosity, could be important for cosmology. 

\section{Conclusions}
\label{conclusions}  
Unknown peculiar velocities are an additional source of uncertainty on the Hubble diagram. Usually, they are taken into account by assuming 150 -- 300 km~s$^{-1}$ as an additional statistical uncertainty in the calculations or by applying corrections based on linear flow maps. However, the velocity dispersion for galaxies inside galaxy clusters can be much higher than these methods account for. In this paper we developed a method for assigning SNe~Ia to clusters, we studied how the peculiar velocities of SNe~Ia in galaxy clusters affect the redshift measurement and propagate through to the distance estimation. We tested the match of \Nsn SNe~Ia observed by \textsc{SNfactory} with known clusters of galaxies and used the cluster redshift to measure the redshifts of the SNe~Ia instead of the host galaxy redshift. Among the full sample of SNe~Ia, \NsnCl were found to be in clusters of galaxies.

The technique we developed improved the redshift measurements for low and intermediate redshifts ($z$ < 0.1) and decreased the spread on the Hubble diagram. When peculiar velocities are taken into account, for the SNe in clusters the \wrms=\wrmsClusDoppler mag is smaller than the \wrms=\wrmsClusNoDoppler mag found when no correction is applied. The correction is statistically significant with a value of \pearsonsigCluster; however, when we exclude SN2006X the significance of the correction decreases to \pearsonsigClusterwithoutSNX.

We also found that the Hubble diagram dispersion of the \NsnCl SNe~Ia that belong to clusters is smaller than for SNe in the field, but with a p-value of \pvalueDispCluster, which is not statistically significant.  Among \NsnCl SNe found in clusters the SNe~Ia hosted by blue spiral galaxies, with high local sSFR, smaller $M_{stellar}$, $r/R200\simeq0.7$ show higher peculiar velocity corrections (see Fig.~\ref{fig:M200_vs_cdeltaz}).

Since the majority of galaxies in the Universe are not found in galaxy clusters, but in filamentary structures such as the Great Wall \citep{Geller89}, SNe~Ia in galaxy clusters are rare in untargeted searches such as \textsc{SNfactory}. Next decade surveys such as ZTF or LSST~\citep{ZTF14,LSST2009} will observe thousands of SNe~Ia and therefore have much larger samples of SNe~Ia in clusters. These can be used to study dependencies between SNe~Ia and host clusters with greater certainty. LSST will be much deeper than \textsc{SNfactory} or ZTF, so the method of cluster selection based only on the presence of X-rays will not be viable until much deeper all-sky X-ray surveys are performed. Even though the impact of peculiar velocities decreases with distance and becomes negligible at high redshifts, SN~Ia rates in clusters \citep{Sharon2010,Barbary12} and the difference in SN light curve parameters inside and outside the clusters could be fruitful avenues of investigation for future cosmological analyses.

\begin{acknowledgements}
We thank the technical staff of the University of Hawaii 2.2-m telescope, and Dan Birchall for observing assistance. We recognize the significant cultural role of Mauna Kea within the indigenous Hawaiian community, and we appreciate the opportunity to conduct observations from this revered site. This work was supported in part by the Director, Office of Science, Office of High Energy Physics of the U.S. Department of Energy under Contract No. DE-AC025CH11231. Support in France was provided by CNRS/IN2P3, CNRS/INSU, and PNC; LPNHE acknowledges support from LABEX ILP, supported by French state funds managed by the ANR within the Investissements d'Avenir programme under reference ANR-11-IDEX-0004-02. NC is grateful to the LABEX Lyon Institute of Origins (ANR-10-LABX-0066) of the University de Lyon for its financial support within the program "Investissements d'Avenir" (ANR-11-IDEX-0007) of the French government operated by the National Research Agency (ANR). Support in Germany was provided by DFG through TRR33 "The Dark Universe" and by DLR through grants FKZ 50OR1503 and FKZ 50OR1602. In China support was provided by Tsinghua University 985 grant and NSFC grant No 11173017. Some results were obtained using resources and support from the National Energy Research Scientific Computing Center, supported by the Director, Office of Science, Office of Advanced Scientific Computing Research of the U.S. Department of Energy under Contract No. DE-AC02- 05CH11231. We thank the Gordon \& Betty Moore Foundation for their continuing support. Additional support was provided by NASA under the Astrophysics Data Analysis Program grant 15-ADAP15-0256 (PI: Aldering). We also thank the High Performance Research and Education Network (HPWREN), supported by National Science Foundation Grant Nos. 0087344 \& 0426879. This project has received funding from the European Research Council (ERC) under the European Union's Horizon 2020 research and innovation programme (grant agreement No 759194 - USNAC). PFL acknowledges support from the National Science Foundation grant PHY-1404070. MVP acknowledges support from Russian Science Foundation grant 14-12-00146 for the selection of SNe exploded in galaxy clusters.
This research has made use of the NASA/IPAC Extragalactic Database (NED), which is operated by the Jet Propulsion Laboratory, California Institute of Technology, under contract with the National Aeronautics and Space Administration. 
Funding for the Sloan Digital Sky Survey IV has been provided by the Alfred P. Sloan Foundation, the U.S. Department of Energy Office of Science, and the Participating Institutions. SDSS-IV acknowledges support and resources from the Center for High-Performance Computing at the University of Utah. The SDSS web site is www.sdss.org. SDSS-IV is managed by the Astrophysical Research Consortium for the Participating Institutions of the SDSS Collaboration including the Brazilian Participation Group, the Carnegie Institution for Science, Carnegie Mellon University, the Chilean Participation Group, the French Participation Group, Harvard-Smithsonian Center for Astrophysics, Instituto de Astrof\'isica de Canarias, The Johns Hopkins University, Kavli Institute for the Physics and Mathematics of the Universe (IPMU) / University of Tokyo, Lawrence Berkeley National Laboratory, Leibniz Institut f\"ur Astrophysik Potsdam (AIP), Max-Planck-Institut f\"ur Astronomie (MPIA Heidelberg), Max-Planck-Institut f\"ur Astrophysik (MPA Garching), Max-Planck-Institut f\"ur Extraterrestrische Physik (MPE), National Astronomical Observatories of China, New Mexico State University, New York University, University of Notre Dame, Observat\'ario Nacional / MCTI, The Ohio State University, Pennsylvania State University, Shanghai Astronomical Observatory, United Kingdom Participation Group, Universidad Nacional Aut\'onoma de M\'exico, University of Arizona, University of Colorado Boulder, University of Oxford, University of Portsmouth, University of Utah, University of Virginia, University of Washington, University of Wisconsin, Vanderbilt University, and Yale University. 
This research has made use of the \textsc{SIMBAD} database, operated at CDS, Strasbourg, France. 
We acknowledge the usage of the \textsc{HyperLeda} database (http://leda.univ-lyon1.fr). 
We have made use of the ROSAT Data Archive of the Max-Planck-Institut für extraterrestrische Physik (MPE) at Garching, Germany. 
\end{acknowledgements}

\bibliographystyle{aa}
\bibliography{Biblio}

\end{document}